'Bridging the gap between emotion and joint action'

*In press - Neuroscience and Biobehavioral Reviews*

Bieńkiewicz, M. M. N.[1], Smykovskyi, A.[1], Olugbade, T.[2], Janaqi, S.[1], Camurri, A.[3], Bianchi-Berthouze, N.[2], Björkman, M.[4], & Bardy, B. G.[1]

[1]EuroMov Digital Health in Motion, Univ. Montpellier IMT Mines Ales, Montpellier, France

[2]UCL, University College of London, UK

[3]UNIGE, InfoMus, Casa Paganini, Italy

[4]KTH, KTH Royal Institute of Technology, Sweden

Corresponding authors:

Marta Bieńkiewicz (marta.bienkiewicz@umontpellier.fr)

Benoît Bardy (benoit.bardy@umontpellier.fr)

EUROMOV, UNIVERSITÉ DE MONTPELLIER,

700, Avenue du Pic Saint Loup,

34090 Montpellier, FRANCE

Funding statement

This work is financed by 'EnTimeMent' project funded by EU Horizon 2020 FET PROACTIVE project (No 824160).

*Bridging the gap between emotion and joint action.* | Bieńkiewicz et al. (in press, NBR)

# 1. General introduction

We thrive on being surrounded by others; we wither when isolated (Baumeister and Leary, 1995). The quantity and the quality of our social interactions are one of the most robust predictors to both how well and how long we live, beating the predictive power of exercise or obesity (Holt-Lunstad et al., 2010). Our brains have been carved by evolution to act together with others towards long-term mutual goals, by emergence of 'self'-transcendental emotions as opposed to immediate and egoistic benefits (Stellar et al., 2017). These emotions (i.e., compassion, gratitude, or awe) exclusively promote coalitional behavior such as caretaking, cooperation and coordination. For these reasons, moving in unison with others is the firmest of social ties, a superglue, pushing us as a group towards more ambitious goals and performance outcomes as opposed to acting as individual units (Duranton and Gaunet, 2016; Hasson et al., 2012; Marsh et al., 2009; Salmela and Nagatsu, 2017; von Zimmermann and Richardson, 2016). Army drills based on marching together bring a feeling of affiliation and facilitate group performance (McNeill, 1995), just like the Haka dance in a rugby match pumps up the morale before meeting the enemy and perceptually diminishes the strength of the rival (Clément, 2017; Fessler and Holbrook, 2014). We appraise being a member of a larger group ('tribe') by aligning our actions with others (Tsai et al., 2011) to show we like them, we care about them and we are ready to work with them (Mogan et al., 2017; Parkinson, 2020). According to anthropological and behavioral research, a group that chants and dances well together also hunts well (von Zimmermann and Richardson, 2016). Throughout centuries, political and religious power holders have engaged crowds during rallies by using repetitive gestures or vocal expressions (Heinskou and Liebst, 2016; Lukes, 1975). Such rituals (for instance the Nazi salute during May Day rally, cf., Allert, 2009) put normative pressure on individuals with the purpose of bringing up a certain collective thought and feeling, captured by a classical, yet poorly understood in neuroscience, sociological concept of 'collective effervescence' (Liebst, 2019; Pickering, 2020). The social morphology of being in a crowd is viewed as essential for motor and emotional synchrony through entrainment of rituals (Borch, 2015; Collins, 2004) with the emergence of contagion hot spots ('transmitters') being the essence of collective effervescence (Liebst, 2019; Zheng et al., 2020).

Surprisingly, although our social interactions are highly intertwined with emotions we convey or receive, emotions and joint actions are primarily analyzed and modeled by different branches of science, and are usually described separately from each other (Salmela and Nagatsu, 2017). Most models of emotion are individualistic and do not explicitly consider the social context of interacting with others, recently identified as the 'dark matter' of modern neuroscience (Schilbach et al., 2013). Exceptions exist in research on empathy (e.g., the

*Bridging the gap between emotion and joint action.* | Bieńkiewicz et al. (in press, NBR)

model of therapist - patient relation by Koole and Tschacher (2016)), and in research dedicated to movement expression and propagation in arts, such as musical ensembles or dance performance (e.g., Alborno et al., 2015; Basso et al., 2021; Camurri et al., 2011; Chabin et al., 2020; Jola et al., 2013). However, the "acting together" component has not yet been tackled (Butler, 2017; Goldenberg et al., 2016; Mayo and Gordon, 2020).

Reciprocally, human synchronization models, which dominate the joint action research, are not usually inclusive of the emotional state of agents and the manifestations of emotions or other affective components in the way agents move to achieve an optimal outcome (Wallot et al., 2016a; Wolpert et al., 2003). This urgency to address emotional social interaction aspects was recently recognized by Shamay-Tsoory et al. (2019) in their social alignment model, which incorporates the emotional component of group behavior. The scarcity of research is quite surprising, taking into consideration that both self-transcendent emotions and cooperation evolved together as strong features of humanity and contributed greatly to human dominance in the animal kingdom (McNeill, 1995).

In this paper, we zoom into a large body of literature in order to give a synthetic review of the current state-of-the-art on emotion and on joint action, currently two separate strands of research[1]. We present evidence showing how these fields are intertwined, and in need of marriage to create a more interdisciplinary and ecologically relevant branch of science. Understanding how we feel impacts the way we act together, and how we act together impacts our emotions, is a societal and scientific challenge of the utmost importance (Barrett, 2017a; Feldman Barrett and Finlay, 2018; Salmela and Nagatsu, 2017; Shamay-Tsoory et al., 2019; Wallot et al., 2016b; Wolpert et al., 2003). As posited by Salmela and Nagatsu (2017), we argue that the emotional component can reveal a lot about the prediction of actions of others, but also about the unfolding of joint acts and their outcomes. We propose a pathway for updates of current models on joint action, such as synchronization, that could be a start to a more integrative approach. We intend to encourage the scientific community to join this conversation, by prompting potential research questions, revise current models of emotion and joint action across neuroscience, computer sciences and social sciences, and to move towards more integrative, social approaches (e.g., Hasson et al., 2012; Hoemann and Feldman Barrett, 2019).

Throughout our review, and in the future directions Section 5, we show evidence that witnesses how the stem of this new research avenue is now shaping the future of human machine interfaces (e.g., robotics, interactive art systems, embodied social media). We believe this new avenue will lend itself to informing occupational health; in promoting efficient and human-friendly working environments and workflows (be it digital or physical) - on a micro-scale, but also crowd management during sport, public and emergency events - on a macro-scale.

---

[1] We ran a Google search and a PubMed search exploiting core keywords linked to emotion and acting together (for the exact terminology used for the search dated 18/05/2020 and updated 04/04/2021 please see Annex 1) and followed up on cross-references.

# 2. Joint action: Humans act together

## A. What is joint action?

Joint action can be regarded as any form of socio-motor interaction whereby two or more individuals coordinate their actions in space and time to bring about a change in the environment. Joint action depends on the following mechanisms: joint attention, representations of others, action prediction and coordination, as well as awareness of oneself and the outcome of the actions of others (Sebanz et al., 2006). Acting together can be emergent or planned and it encompasses the level of intentions, action plans and movements (Knoblich et al., 2011). There is a great variety of terms used by scientific communities in reference to the phenomenon itself: joint action (Sebanz et al., 2006), interpersonal coordination (Mayo and Gordon, 2020; Vicaria and Dickens, 2016), interpersonal adaptation (Burgoon et al., 1995), nonverbal adaptation (Bodie et al., 2016) or even social interactions (Hasson and Frith, 2016). Furthermore, scholars have often offered narrow typologies for joint action subclasses such as physiological or behavioral synchrony and behavioral matching (Mayo and Gordon, 2020), interactional synchrony and behavioral matching (Bernieri and Rosenthal, 1991), mimicry and imitation (Chartrand and Bargh, 1999). Differences in the terminology used by researchers usually stem from stressing and developing models of a particular aspect of socio-motor interaction. For example, Jarrassé et al. (2012) distinguished divisible, interactive and antagonistic tasks, focusing on the objectives and roles each interactant follows (see Table 1). Similarly, Clark (1996) and Toma (2014) both accounted for the temporal relationship of interactions and concentrated on perceptuomotor aspects of interaction during communication. Burgoon et al. (1995) proposed a more holistic framework of different dimensions (e.g., directedness, timing, measure of behavioral change, intentionality, etc.) to study dyadic adaptation (see Fig. 1).

*Bridging the gap between emotion and joint action.* | Bieńkiewicz et al. (in press, NBR)

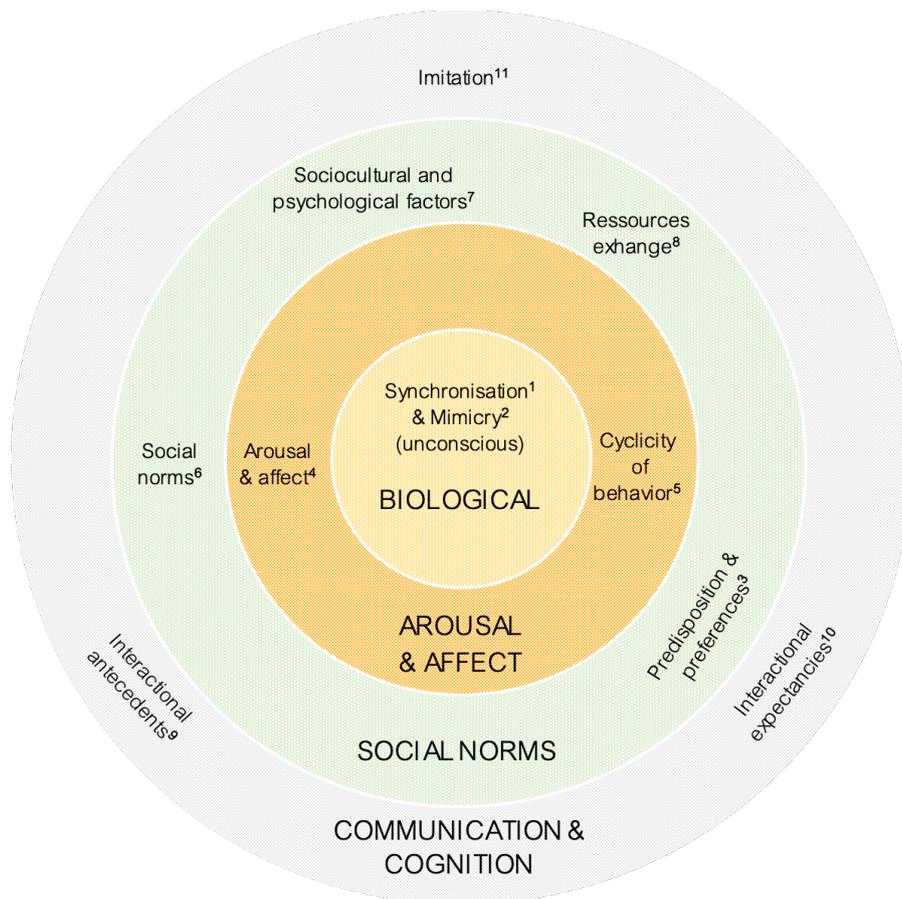

Fig. 1. Levels of joint action (after Burgoon et al.,1995); 1. (Bernieri and Rosenthal, 1991); 2. (Chartrand and Bargh, 1999); 3. (Argyle and Cook, 1976); 4. (Cappella and Greene, 1982); 5. (Altman et al., 1981); 6. (Gouldner, 1960); 7. (Giles et al., 1973); 8. (Roloff et al., 1988); 9. (Patterson, 1983); 10. (Hale and Burgoon, 1984); 11. (Bavelas et al., 1986).

What remains uncertain in this classification is the required number of characteristics that need to be satisfied for interaction to be regarded as joint action, and its sub-level type, and whether the individual change occurs prior, during or after the interaction (Burgoon et al., 1995). Importantly, joint action does not automatically imply cooperation; it can also indicate competition in terms of individual performance within the group, depending on whether action goal is driven by individual gain (be better as a unit than others) versus collective gain (be better as a group than others) (Tuomela, 2011), leading to multiple possible patterns of coordinated dyadic and group joint action. Table 1 presents an overview of the major sub-types of socio-motor interaction discussed further in this review.

*Bridging the gap between emotion and joint action.* | Bieńkiewicz et al. (in press, NBR)

Table 1. Summary of distinct features of different patterns (subtypes) of socio-motor interaction extracted on the basis of literature review in Section 2.

| ARCH TERM | SOCIO-MOTOR INTERACTION (AGENTS SHARING THE SAME SPACE AND TIME) | | | |
|---|---|---|---|---|
| SUBTERM | MIMICRY 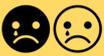 | JOINT ACTION | | |
| | | IMITATION 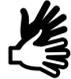 | SYNCHRONY 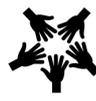 | COOPERATION 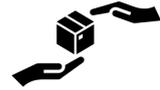 |
| SUBCLASS | physiological, behavioural | automatic (spontaneous), intentional | | interactive, divisible, antagonistic |
| METRICS | muscle, physiological, neural activity | body position, neural activity | phase, frequency, time in/out of sync, coupling, re-occurrence | action goals, effort, phase, trajectories, response time |
| SPACE RELATIONSHIP | copy | | copy or not, temporal alignment pivotal | connected, but not a copy |
| TIME RELATIONSHIP | DELAYED | | ALIGNED | |
| PREDICTION/ ANTICIPATION | NO | | YES | |
| AWARENESS | LOW/ SPONTANEOUS | - - - - - - - - - - - - - - - - -> | | HIGH/ INTENTIONAL |
| TIMESCALES | SHORTER PHASIC | - - - - - - - - - - - - - - - - -> | | LONGER/ CHRONIC |
| SOCIAL CONSEQUENCES | rapport, empathy, emotion recognition | affiliation, learning, empathy, culture | affiliation, interpersonal attractiveness, cohesion | higher order, complex, longer perspective societal goals |
| EXAMPLES | facial expression; heart rate | body alignment | hand clapping, haka, brain to brain coupling | orchestra, line manufacturing, see-saw |
| EMERGENCE DURING LIFESPAN | 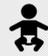 | 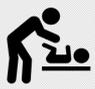 | 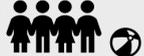 | 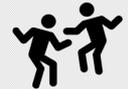 |

*Note.* Selected features (first column on the left) were identified as informative for understanding the subtle differences between different modes (patterns) of socio-motor interaction. Icons at the bottom depict approximations of developmental stages: babies; toddlers; preschool and school age; teens and young adults.

*Bridging the gap between emotion and joint action.* | Bieńkiewicz et al. (in press, NBR)

## B. Models of human group synchronization

As a specific branch of joint action research, the synchronization of a group of agents - such as humans and other animals – or by robotic or digital agents, all underlying the achievement of a common goal, is a robust example of dismissing the emotion component during socio-motor interaction. The state-of the-art in this domain spans over several scientific disciplines including ethology (Couzin et al., 2005) cognitive and movement neurosciences (e.g., Alderisio et al., 2016), robotics (Iqbal and Riek, 2019) and various branches of mathematics and physics (e.g., Ott and Antonsen, 2017; Strogatz, 2004). Synchronization phenomena have been investigated between individuals, ranging from $N=2$ (e.g., Noy et al. (2011) to 7-10 (e.g., Alderisio et al. (2017)) to $N>10$, such as in human crowds, (e.g., Gallup et al. (2012); Rio et al. (2018)). Simply stated, synchronization from a physical principle requires two conditions to be met, (1) a certain behavioral proximity of the systems to be synchronized, such as a common movement amplitude or frequency, and (2) a coupling function between them, through for instance informational exchanges. Typically, metrics for synchronization built on frequency (e.g., the relation between individual and group frequencies), phase (e.g., the order parameter of the synchronization), and their stability characteristics (e.g., Pikovsky et al. (2002), allow us to capture socio-motor coordination characterized by periods of synchronization and desynchronization (Feniger-Schaal et al., 2018; Mayo and Gordon, 2020; Noy et al., 2011).

The generalization of synchronization principles to situations involving more than two agents remains a very recent enterprise[2]. In a nutshell, models of human synchronized behaviors can be categorized in top-down estimation models and in bottom-up self-organized models. In the first category (e.g., Takagi et al. (2019), efficient synchronization between connected participants sharing a common goal is ensured by inference of the shared intention from perceived collective information and consequent adaptation of each individual motion planning. The second category concerns models proposing that synchronized motion observed at the collective level emerges from local interactions between nearby individuals. These models, such as the Kuramoto model (e.g., Strogatz (2004)) or the extended HKB model (see Kelso (2021) for a review) aim to decipher how local informational exchanges and motor adaptations contribute to that emergence. For rhythmic biological movements, coupled oscillator dynamic models have begun to explore perceptuo-motor synchronization phenomena in situations where $N>2$ (e.g., (Alderisio et al., 2017; Bardy et al., 2020; Zhang et al., 2019).

---

[2] In this article, we are not reviewing experiments and models of collective motion in the animal kingdom such as bird flocks, fish schools, or fireflies' synchronized flashes, although some of them have influenced emergent models of human synchronization (see (Clark, 1997; Frijda, 2007)).

It is striking to observe that a single system of differential equations (Kuramoto or extended HKB model) can capture the complexity of our interactions in perceptuo-motor tasks, including leadership properties (Alborno et al., 2015; Aucouturier and Canonne, 2017; Gnecco et al., 2013; Hilt et al., 2019; Varni et al., 2010), paving the way for the incorporation of yet missing emotional components (expressive qualities of movement). Anticipating this trend, Varni et al. (2019) recently proposed a Kuramoto-based model of entrainment in music performance of an orchestra, with two components (cf. Phillips-Silver and Keller (2012)): temporal and affective. While the first type relates to rigid temporal alignments in synchronization, the second type, initially developed in childhood during socio-motor interactions with primary carers (Barrett et al., 2007), allows for sharing emotional qualities. Thus, in musical ensembles, entrainment is seen as a temporally flexible organization between musicians, with successive periods of synchronization — stronger at the beginning and at the end of each musical phrase compared to the middle portion (Yoshida, 2002) — creating space for the unfolding of expressive performance cues.

The models of perceptuo-motor social synchronization reviewed above are only a sample of a wide state-of-the-art across several fields of research, and additional branches, for instance social contagion models (Farkas et al., 2002; Mann et al., 2013; Ugander et al., 2012), have been omitted. However, it is striking that *none* of these models have addressed the various emotional qualities that are intrinsic to our efficient and socially relevant cooperative actions. Emotional qualities have occasionally been manipulated (Zhao et al., 2015) or measured (Zhang et al., 2016) during dyadic interaction, although not to our knowledge when $N>2$, and have not been considered in the various formalisms. In Subsection 5B, we will propose new directions to incorporate these emotional qualities in joint action models of social synchronization.

## C. Neural origins of acting together

Moving together with others, as discussed in the previous section, results from the coupling between dynamical systems sharing the same space at the same time. Any stable and adaptive behavior emerges as the resultant interaction of internal dynamics and the physical and informational structure of the environment in which the dynamical systems evolve (Gibson, 1979; Warren, 2006). Agents learn how to act in the most efficient manner, by searching, for instance, for the most stable attractors (Thelen and Smith, 1994), or by exploiting optimal control (Todorov and Jordan, 2002). This internal self-organization of agent systems in response to external demands is fueled by previous experiences and genetic make-up (e.g., Hainsworth, 1989). Perception naturally plays a key role in delivering information to the brain about the environment and enabling us to efficiently maneuver in it. The investigations into

*Bridging the gap between emotion and joint action.* | Bieńkiewicz et al. (in press, NBR)

further neural underpinnings of human joint action can be divided into two main areas of research that we herein review in turn: (i) the ability to imitate and understand movement intention in others, and (ii) the synchronization of brain activities during multi-agent scenarios.

Understanding movement intention in others: The cognitive ability to extract meaning from the perceived actions of others shapes how we cooperate and communicate with them (Rizzolatti and Craighero, 2004). Primates and other social animals, such as birds, are hard-wired to copy the behaviors of group mates (Heyes, 2021). Invasive brain recordings in macaque monkeys identified the neural cornerstone of this ability to visuomotor neurons in the premotor cortex V5. Commonly referred to as mirror neurons, they discharge both when a monkey observes another monkey or another human performing an action, as well as when the monkey performs the action itself (i.e., eating a peanut). Direct single cell recordings are rare in human research, but scarce evidence suggests the existence of similar neurons in the human brain. The supplementary motor area and hippocampus were identified as active during both action observation and action execution, such as reaching (Iacoboni et al., 2005). Interestingly, parts of this network were shown to be active only during observation, indicating an inhibitory role for releasing the imitation response, and other areas discharged only during execution, suggesting a dissection of this network, referred to as the mirror neuron system (MNS) for action observation (perception) and execution (action). Here, it needs to be noted that the mirror neurons represent an undefined percentage of all neurons in the network for action observation and execution (di Pellegrino et al., 1992), and consequently the validity of using the term MNS itself can be contested; Mirror neurons contribute to, but do not dominate, complex neural networks involved in high-order understanding of action (Heyes and Catmur, 2021). There is an emerging consensus that mirror neurons play a key part only in the processing of low-level features of action for recognition and discrimination, such as body movement topography of another agent, but not of higher-level features such as intention or goal reading (Thompson et al., 2019). Ramachandran (2000) already suggested that the human leap in evolution – the creation of unique aspects of human culture such as language and arts – is endowed to the MNS, as it fosters knowledge transfer between individuals through imitation. Very recently, Heyes (2021) pointed out that imitation is shaped by development and culture, as newborns do not spontaneously imitate but learn to do so via sensori-motor learning, which promotes transformation of neurons to mirror neurons. Emotion and the reward-processing circuitry (the medial orbitofrontal cortex/ventromedial prefrontal cortex) were shown to be linked to the observation of being imitated by another (Kühn et al., 2010), an important hint for the composition of the social glue that acting together promotes.

Some authors (Becchio et al., 2012; Centelles et al., 2011) attribute to MNS, in conjugation with the mentalizing network, the ability to distinguish individual movements from those that are socially connected, and to understand the personal relevance of the movements of others (Kourtis et al., 2010). Exaggeration or modulation of kinematics in order to convey (socially relevant) intention is often referred to as sensori-motor communication (Pezzulo et al., 2013). For example, kinematic differences in the proximal and distal parts of body movements provide pivotal information about whether someone is reaching for a cup to take a sip or to pass it on, and allow us to make adequate predictions and react accordingly before the end of the movement (Ansuini et al., 2014; Cavallo et al., 2016; Soriano et al., 2018). Regardless of the not-yet-fully-understood contribution of mirror neurons themselves in the network, the MNS acts like a bridge between first-and-third-person experience, allowing for replay of cognitive representation, but safe-guarding the correct placement of the self (Gallese et al., 2004; Meltzoff and Decety, 2003).

Synchronization between brains in shared space: The moment one person interacts with another person in a shared space, we can no longer analyze those entities separately, not only in terms of motor output, but also in terms of brain activity (Balconi and Vanutelli, 2017). Eye-to-eye contact engages language production and reception areas inviting social expression and engagement (Hirsch et al., 2017). Over the last two decades, the state-of-the-art in social neurosciences has indeed shown considerable evidence that our brain is entrained by the structure of physical interaction in the same way it is entrained by the activity of another brain when such interaction is of social nature (Hasson et al., 2012). During social interactions our brain activity is coupled with that of others (Hari and Kujala, 2009), by viewing the same content (Hasson et al., 2004; Nummenmaa et al., 2012), by movements (Basso et al., 2021; Dumas et al., 2010), and even by speech (Hasson et al., 2012; Jasmin et al., 2016). Discussed techniques, referred to as 'hyperscanning' (as more than one brain is being recorded) were specifically employed to look at dyads imitating meaningless gestures, and they identified the alpha-mu band as critical for the coordination of interpersonal dynamics, with asymmetrical patterns in brain activity reflecting the imitator-model roles (Dumas et al., 2010). In another context, two particular indexes were identified within this band in the centroparietal brain area - $phi_1$, $phi_2$; and were linked to the coordination of individualistic versus cooperative behaviors in dyads, translating into inhibition and enhancement of MNS activity (Tognoli et al., 2007). Importantly, the multi-person framework of EEG research has also started to address the niche of social affective interaction (Acquadro et al., 2016). For instance, more pronounced brain-to-brain synchrony (measured with EEG) was found in school classmates (*N*>2) sharing attention and engaging in face-to-face interactions (Dikker et al., 2017). Also, Babiloni et al. (2012) found that people with high empathic disposition in saxophone quartets (*N*>2) had higher alpha

*Bridging the gap between emotion and joint action.* | Bieńkiewicz et al. (in press, NBR)

desynchronization in the BA 44/45 Broadmann's area during the observation of video recordings of their performance as a musical ensemble. We view this and other recent studies (Chabin et al., 2020; Czeszumski et al., 2020) as promising attempts to investigate the neurophysiology of embodied emotions during joint action.

## D. Social benefits of acting together

Acting together has been shown to have profound psychosocial consequences, with evidence coming from studies looking primarily at dyads and to lesser extent, groups. The story of the interplay of emotion and synchrony starts in the most primal context of interpersonal relationships – the dyad composed of an infant and a mother. Human infants have no capacity to survive on their own and need a primary carer to regulate their physiological balance (allostasis) (Atzil and Gendron, 2017; McEwen, 2000; Van Der Veer, 1996). Emotionally receptive parents cradle their infants on the left side of their body allowing the flow of visual and auditory information to travel directly to the right hemisphere empathy circuits (Malatesta et al., 2019). Multi-modal channels of bidirectional, physiological concordance with caregivers were identified in infants as early as three months old via heartbeat rate, pupil size mimicry, vocal and affective non-verbal expression (Aktar et al., 2020; Feldman et al., 2011; Palumbo et al., 2017). Further, physiological and movement couplings were found to emerge only when the infant is unsettled, resulting in the parent reducing their own arousal to stabilize the infant (Wass et al., 2019). These first experiences of synchrony were identified to be a cornerstone for healthy emotional development (Feldman, 2012). Parental mirroring allows children to learn their own emotional responses, recognize and label them (Atzil and Gendron, 2017; Pratt et al., 2017). Motor synchrony fosters the building of first trust relationships and prosocial behaviors in human development as early as 14 months, when infants are more likely to pick up a toy dropped by a stranger who bounced with them in synchrony a moment before (Cirelli et al., 2014). Later in development, from toddlerhood to teenage years, other aspects of social cognition and self-regulation are carved by play experiences with caregivers and peers using those very foundations (Nijhof et al., 2018; Viana et al., 2020; Williams et al., 2020). Play promotes learning by imitation in the animal kingdom, through discovering how to act with others, and understanding what others feel (Feldman, 2007). This is a bedrock for empathy and social interactions in humans (Donohue et al., 2020; Viana et al., 2020; Xavier et al., 2016), and perhaps is one of the reasons why ability to synchronize movements with others continues to develop until adulthood (Su et al., 2020b).

Moving in unison thus acts as a social glue and reinforces cooperation (Hoehl et al., 2021; Wiltermuth and Heath, 2009), with a sense of affiliation between strangers (Cacioppo et al.,

*Bridging the gap between emotion and joint action.* | Bieńkiewicz et al. (in press, NBR)

2014; Kragness and Cirelli, 2021) and boost in self-esteem (Lumsden et al., 2014), arising even from very simple movements such as synchronous finger tapping (Hove and Risen, 2009). The social cohesion instigated by this motor synchronization (Lakens and Stel, 2011) is part of a virtuous cycle and improves actual performance on subsequent joint action, by increasing the perceptual sensitivity of agents towards changes in the environment, those related for instance to the movements of others (Valdesolo et al., 2010).

In that way, motor synchronization seems to be a currency for our social likes and dislikes. Interpersonal attractiveness and likeability of an interaction partner is linked to the magnitude of effort in coordinating with them (in terms for instance of the relative phasing of our body movements). When engaged in synchronization, we try harder with the person who seems happy and non-threatening (Zhao et al., 2020), or who we find attractive (Zhao et al., 2015). If an interaction partner makes a bad first impression, the chances are we will not put in our best to align our actions with them, as demonstrated by Miles et al. (2009) in a study looking at dyadic synchronous stepping. This relation is bidirectional, as moving in unison fosters interpersonal attractiveness. For instance, Cheng et al. (2020) found that during paired walking tasks a phase synchronization time ratio (how much time people walked in phase with each other) was predicted by how much they liked a stranger based on their initial impression. After a period of silent and 'chatty' walking, strangers reported increased affiliation with the person as a consequence of synchronous walking. Atherton et al. (2019) reported decrease in prejudice towards another ethnic group after physical and imagined walking.

Even if people are not 'on the move' across space, their bodies show gradual convergence towards postural alignment over the course of almost any dyadic interaction (Chartrand and Lakin, 2013; Fujiwara et al., 2019; Paxton and Dale, 2013). In human psychotherapy, which aims at aiding any shortcomings in emotional regulation in adulthood, therapists usually build a 'trust' relation (therapeutic alliance) via the above-mentioned elements of speech synchrony, as well as body movements (Bar-Kalifa et al., 2019; Lutz et al., 2020), to access implicit and explicit emotional regulation of the patient (Koole and Tschacher, 2016). For instance, head position synchrony between therapist and patient has been found to be linked to the overall therapy outcome, whereas body alignments were noted on shorter timescales to be predictive for each session outcome (Ramseyer and Tschacher, 2014) as well as experience of pain with a therapist (Goldstein et al., 2020). Weaker coordination patterns of head motion (angular displacement and velocity of the head's yaw and pitch) were also reported during conversations involving arguments between romantic partners (Hammal et al., 2014) and were proposed to be predictors of the quality of rapport with a therapist (Goldstein et al., 2020). Interestingly, people who live together, and who are hence in permanent dyadic interaction,

*Bridging the gap between emotion and joint action.* | Bieńkiewicz et al. (in press, NBR)

such as roommates or couples, become alike, over time, in terms of their emotional reactivity and emotional expressions (Anderson et al., 2003). During conversations adults show convergence in use of certain speech elements such as figures and grammar (Hasson et al., 2012). Joint speech research revealed that merging ourselves with others is visible in the neural activation patterns that are different from speech production alone (Jasmin et al., 2016), putting joint speech in the realm of dynamic interplay of socially shared cognition (Cummins, 2014).

In a study on large groups (*N*>5) von Zimmermann and Richardson (2016) demonstrated that synchronous vocalization with others enhances memory performance and group effort, providing evidence for hidden wisdom of group rituals such as dancing, or singing, or when marching to fight a rival. Similarly, physiological concordance emerging between newly met group members (i.e. intervals between heart beats) explained one sixth of the variance of the performance in a drumming task in another group study (Mayo and Gordon, 2020). Both physiological and motor synchronization levels were predictive for the subjective sensation of group cohesion. Also, a study by Mønster et al. (2016) looked into the synchronization of bio-signals (heart rate and electromyography of face muscles) during cooperation (line-manufacturing of paper boats by triads). Participants were induced emotionally by the researcher acting either warmly or coldly in the interaction with the group. Exchange of smiles was linked to group cohesion (primed with the warm behavior from the experimenter), whereas synchronous increased skin conduction (interpreted as experiencing tension in a group caused by coldness in the demeanor of the experimenter) correlated negatively with measures of cohesion.

In sum, the motor and physiological coupling between humans is hardwired, and develops through childhood, to deploy group affiliation (we are members of the same tribe) and maximize collaborative efforts. Research developments in dyad research have provided ample evidence for the social benefits of interpersonal alignment, with group research still being under-addressed. Therefore, the future of neuroscience of social interactions needs to include not only a second-person perspective (Schilbach et al., 2013) but also a multi-person or multi-agent perspective, and it constitutes the *raison d'être* of our review.

## 3. Emotions

The previous section synthesized the current state of the art in the main branches of sciences investigating how we act together, from physics-based models of synchronization to social neuroscience in humans and other animals, embodied cognition and developmental psychology. While we demonstrated that emotion is not a dominant focus to understand how people move in a group, we also mentioned some burgeoning research incorporating emotional qualities and evaluating how they affect embodied social interactions. In the current section, we reciprocate with a first analysis of current models of emotion, how they largely ignore joint action and the above models, with some recent exceptions paving the way for a more integrated approach.

## A. What is an emotion?

Since James' attempt to answer the question "What is an *emotion?"* (1884), an unresolved debate started on the inherently elusive nature of the phenomenon (Scherer, 2005), highlighted by Fehr and Russell's (1984) remark that "everyone knows what an emotion is, until they are asked to give a definition" (p. 464). Four decades ago, Kleinginna and Kleinginna (1981) found 92 different definitions of emotion and stressed the need for consensus. Despite substantial advancements, there is not a unified theory of emotion that would exhaustively address all the fundamental questions (Reisenzein, 2015). Interestingly, the etymology of the word "emotion" contains in itself "motion" or *emovere*, in Latin "to move", and the word "affect", more general, which relates to "producing changes" (2020[3]). Arguably, a straightforward example of emotion as a driving force of change comes from the so-called fight-or-flight response (Cannon, 1953). When scared, we prepare to run away in order to withdraw ourselves from the perceived source of danger, and when angry, we prepare to stand up and fight against the threat (Cannon, 1953; Jansen et al., 1995). Stemming from this perspective, emotional arousal holds a motivational function (Reisenzein, 2015), previously conceptualized as a mode of 'action readiness' ('Ur-affekte'; Kafka (1950)). Emotions allow us to adapt to a given set of circumstances with the aim of survival and enhancing wellbeing (maintaining allostasis), thus they entrain different action tendencies to satisfy different needs (Frijda, 2007; Frijda et al., 1989; Frijda and Parrott, 2011; Ridderinkhof, 2017), weighed up by cognitive processes against individual cost/gain and previous experience (Ferrari, 2014; Kiverstein and Miller, 2015; Padoa-Schioppa, 2011).

Classically, there were two main conceptual frameworks for studying emotion: (i) discrete models, representative of individuals emotions (e.g., anger, joy, fear, etc.), pioneered by Darwin (1872) and later developed, for instance, by Izard (1971); and (ii) ls from the environment) and previous experiences feeding in dynamic (visceromotor, motor and sensory) predictions. In this context, emotions are not demarcated events, but derivative of the constant interaction between complex dynamics of the nervous system, the body, and the surrounding environment. We experience the world factually, but it is the visceral reactions that validate the experience as 'real' (Duncan and Barrett, 2007). In a Higher-Order Theory of Emotional Consciousness proposed by LeDoux and Brown (2017), extending the high-order theory of human consciousness to emotions, the sense of self is core to emotional experiences. Although sub-cortical circuitry, such as fear or survival circuit are crucial for providing input for adaptive behavioral responses, they place emphasis on all conscious states (i.e., emotions) being instantiated in the general cortical network of cognition (cortical circuits).

---

[3]    (2020). *Definition of emotion* [online]. Oxford University Press. Available at: https://www.lexico.com/definition/wake (Accessed: 14 November 2020).

Recent neuroimaging studies have shown that the emotional network is widespread in the human brain, and embraces cognitive areas (such as anterior frontal areas) beyond typical affective ones, such as amygdala, diencephalon and brainstem (Duncan and Barrett, 2007; Kiverstein and Miller, 2015) and motor readiness circuitry - premotor area (BA6) (Costa and Crini, 2011). Among other studies, Jastorff et al. (2015) found that distinct categories of emotion emerged only when looking at multi-voxel activity patterns in fMRI during discrimination of visual and dynamic emotional stimuli of different saturation. During the resting state in the MNS network, four hubs were identified as connecting points in the right hemisphere: anterior insula, right anterior cingulate, right precentral sulcus and right fusiform gyrus. Four points of connection with other structures were identified in the emotional network: right amygdala, right insula, left putamen, and left middle STS. Interestingly, Costa et al. (2014) demonstrated, using EEG recordings, that emotions from different categories overlap spatially in activation patterns with the emotional brain network, but also show distinct temporal signatures (i.e. time to peak). This aligns with Barrett (2017a, 2017b) argumentation that there are no specific pathways for emotion categories, but different neural activation can lead to the same emotion ("many-to-one") and reversely the same network can give rise to different emotions ("one-to-many") (i.e. notion of degeneracy (Edelman and Gally, 2001)).

## B. Understanding others' emotions

Despite the recent progress in emotion research, little is known about the actual dynamic link between emotion and social interaction ((Butler, 2017, 2015)), and the models briefly reviewed above remain largely silent about the social nature of emotion. In this section, we focus on the "shareability" of emotions, first through the prism of empathy and mechanism of mimicry, before addressing scarce evidence for the existence of group emotion.

Empathy: Broadly, empathy relates to understanding and 'sharing' what other people feel, need or want to do (Bloom, 2017; Ferrari, 2014), but does not infer action itself. If we refer to empathy as a parameter in prosocial behavior, we mostly mean empathetic distress —

*Bridging the gap between emotion and joint action.* | Bieńkiewicz et al. (in press, NBR)

experiencing discomfort caused by perception of distress in others — is equal to emotional ('hot') empathy (Bloom, 2017). However, other possibilities of sharing an affective state include cognitive ('cold') empathy (conceptually understanding what another person experiences), emotional contagion (i.e. 'catching' anxiety because we share physical space with someone who is anxious), and finally, compassion leading to helping behavior and altruism (Bloom, 2017; Preston and de Waal, 2002).

The well-known Perception-Action model (PAM) of empathy (Preston and de Waal, 2002) proposes that the perception of another agent's affective state activates the same neural representation of the observer (without any particular 'empathy' center), leading to activation of the same somatic and autonomic responses, an idea that was first conceived by Darwin (1872), later reinforced by Hommel (1998, 1997). Research on MNS has revealed that we do internally simulate the actions we perceive (measured as shorter reaction times in consecutive execution of action performed by another agent) (Craighero et al., 1998). For example, whether we observe a facial expression of disgust, or imitate it, the same parts of the brain activate as during actual experience of disgust (Carr et al., 2003). In the same vein, observing fearful bodily expressions in others, activates motor readiness circuitry in the observed (Borgomaneri et al., 2015). Human toddlers show distress when observing others in distress (Zahn-Waxler et al., 1992), and they 'try on' emotional reactions they observe in others to see how they feel, without particular need of their own to be broadcasted (Einon and Potegal, 1994). This behavior in adults, e.g., bursting out crying when someone else does, would be considered as pathological or maladaptive, deprived of emotional containment and dissociation between self and others.

The neural underpinnings of embodied empathy, i.e., the imitation of emotional facial expressions, have so far been pinpointed to the right ventral premotor cortex (Leslie et al., 2004), with the inferior parietal lobule identified as a locus attributing a sense of agency for the self and others (Meltzoff and Decety, 2003). The brain activation of visual information containing key action features flows from the superior temporal cortex to the posterior parietal (to simulate action and code kinematics), involves frontal MNS (to identify action goal), and flows back to the superior temporal cortex to inform it about action prediction and imitation plan if needed (Carr et al., 2003). In this connectivity model the insula relays the action representation to the limbic system from the MNS and motor areas. Although simulation is a key response for understanding what others might feel, people are usually not an echo chamber for other people's feelings. The neural architecture of empathy is complex and sophisticated, endowing a multitude of social interaction scenarios, and is intertwined with cognition (Bernhardt and Singer, 2012; Ferrari, 2014). A double dissociation

*Bridging the gap between emotion and joint action.* | Bieńkiewicz et al. (in press, NBR)

neurophysiological mechanism was proposed based on lesion studies for (i) cognitive (cold) empathy embedded in the ventromedial prefrontal areas, and for (ii) emotional (hot) empathy rooted in the inferior frontal gyrus (Shamay-Tsoory et al., 2009).

<u>The prism of mimicry</u>: The motor phenomenon of mimicry, or 'chameleon effect' (Chartrand and Bargh, 1999), overlapping in some ways but not to be confounded with empathy ('I felt what you feel'), relates to involuntary mirroring of expressions of our interaction partners ('I saw and show what you feel`). Mimicry, unlike imitation and synchronization, is unconscious[4], and at least partially independent of the performance ability in the latter, despite clearly being nested in the same behavioral spectrum with strong functional interconnections (Genschow et al., 2017; Rauchbauer and Grosbras, 2020). This comes with the caveat that mimicry is usually recorded in a naturalistic observation, such as matching up facial expression (rapid facial reaction) during face-to-face contact with another person (Moody et al., 2007), whereas imitation and synchronization are most often elicited and captured in less ecological scenarios (with instruction). Unintentional mimicry of body and vocal expressions unifies emotional state by means of evoking the same internal responses in agents (Hatfield et al., 2011, 1993), and is perhaps the closest to the concept of emotional contagion. Neuroimaging studies (hyperscanning mock setup for fMRI) show a similar time-locked pattern of brain activity between people subjected to watching emotional excerpts from movies (Nummenmaa et al., 2012) and listening to autobiographical stories retold by interaction partner (Smirnov et al., 2019). Adults also align their heart rate variability (Scarpa et al., 2018), pupil diameter, respiration rate, body temperature and electrodermal activity by their mere physical presence in one space (see Palumbo et al. (2017) for a comprehensive review on physiological synchrony). Emotional response to others via mimicry is also mediated by oxytocin hormone (Korb et al., 2016) which, in surplus, can increase mimicry of emotional expressions such as demonstrated in a double-blind study on males looking at adult and infant expressions. Similarly, Festante et al. (2020) looked into EEG after intranasal oxytocin administration and found that the enhanced alpha range of the mu rhythm desynchronization had an impact on the sensorimotor circuits involved in social perception and action understanding. Complimentary to this evidence, The Neurocognitive Model of Emotional Contagion (Prochazkova and Kret, 2017) proposed multi-modal connections between motor mimicry (facial and body, inclusive of eye synchrony) and autonomic mimicry (physiological synchrony). In this model, both mechanisms are considered separately, but operate on the social interaction interface (cognition), with MNS being an engine for shared emotional arousal and steering empathic behavior. Also, the mimicry model by Wood et al. (2016) proposes decomposition of the perception-action loop into sensorimotor layers, encompassing functions such as moderation of emotion, prior beliefs, arousal and adaptive behavioral response. Mimicry, even on a fast timescale under 1000ms (as demonstrated in EMG study of facial muscles), is intermediated by the persons' own affective state and the environmental context (Hess and Fischer, 2013), as well as by affiliative goals (Rauchbauer et al., 2015). Interfering with facial mimicry (i.e., mouthguards) blocks out emotional recognition of the body and the facial expressions of fear and happiness in others (Borgomaneri et al., 2020; Rychlowska et al., 2014). This effect was reported to be mitigated by individual levels of empathy; people with higher empathy levels rely less on mimicry to recognize emotional states in others, suggesting at least partial functional independence of empathy from mimicry (Borgomaneri et al., 2020). The social function of mimicry is also associated with longer timescales. Hogeveen et al. (2015) identified that mimicry might increase social attunement for a longer period than initial interaction via increased mu-suppression activity in MNS.

---

[4]While for some scholars e.g., (Barrett et al., 2019b) mimicry is the process of an unconscious copying of other's postural, facial and other behaviors, for other authors e.g., (Centelles et al., 2011; Clarke et al., 2005; Nackaerts et al., 2012) to mimic is to intentionally imitate the behavior of the other, we us term 'mimicry' in this review in the former definition.

Based on the evidence above, both empathy and mimicry are key players in emotion propagation, interconnected at the functional level and bridged on the neural level by MNS. However, the dynamics of spreading motor, neural and physiological embodiment of emotions between several agents, on short and transient timescales, as well as its impact on joint action, short and long term, remains largely unexplored today. One early exception is the model proposed by Kelly and Barsade (2001), considering group emotion as resulting from individual states of the agents (shared implicitly and explicitly), captured as an affective tone that molds cohesion in the group and fluctuates. For example, adopting a group identity can inflate saliency of negative emotions, such as anger, through linking group-based appraisal to group emotion and prompting pre-designated behavioral response (Kuppens et al., 2013). One important consequence of this line of research is linked to education, as group emotions in classrooms were reported to mediate attention sharing and learning, cornerstone of long-term academic achievements (Eilam, 2019). To understand further the dynamics of 'sharing' emotion between multiple agents as they occupy or move in the same space at the same time, we give an overview of the affective embodiment research on the multi-modal signals that can carry information about emotion qualities of the agents.

## C. The embodiment and automatic recognition of emotions

The role played by the various layers of the moving body as both receptacles and vehicles of emotional experiences has been largely addressed. Here we briefly synthesize research in

*Bridging the gap between emotion and joint action.* | Bieńkiewicz et al. (in press, NBR)

psychology, neuroscience and affective computing revealing how emotional qualities emerge from multi-modal inputs (i.e. Arias et al. (2018)), from face to whole body and physiology, before showing the unanswered questions at the heart of this review.

Face: As the most ancient and still most dominant area of emotion research concerns faces (Ekman, 1992), it is not surprising that the majority of research efforts in affective computing have been targeted to facial expressions (de Gelder, 2009). For instance, the Facial Action Coding System (FACS) was developed to provide an objective, standardized, and measurable coding system of emotional facial expressions (Ekman and Friesen, 1978). Combination of facial muscle activations (e.g., raised eyebrows, wrinkled nose, tightened lips) are differentiated as Action Units (AU) of micro expressions — i.e., instantaneous facial movements hardly perceived by the naked eye — and are subsequently identifiable as an experienced emotion (Ekman and Rosenberg, 2005). For instance, lip-corner raising is identified as an AU12 that, among others, is associated with joy. Thanks to progress in computational techniques and the increase in size of facial expression datasets, raw data rather than FACS or hybrid machine learning architectures are now used to let the mathematical models identify the relevant muscle patterns. For a detailed overview on the embodiment of emotions in facial expressions please refer to Barrett et al. (2019). For an overview on automatic recognition of facial expression see (Küntzler et al., 2021).

Whole body: A growing body of literature has shown that body expressions are at least as powerful as facial expressions in conveying affect (Atkinson et al., 2004; Hadjikhani and Thilenius, 2005; Wallbott, 1998). Studies have shown that in certain situations or for certain emotional states, the body is more informative than the face. For example, In the case of incongruence between facial and body expressions, the studies (Meeren et al., 2005; Van den Stock et al., 2007) show that body posture has a strong influence on the perceived emotion. These findings were also supported by Aviezer et al. (2012). They show that evaluations made on body expressions rather than on facial expressions lead to more accurate assessment of the affective valence of the situation that triggered such expressions. De Gelder (2009) added that the body does not only convey a person's affective emotional state but also her actions and intention in response to it. Further, it should be considered that at close distance people may possibly rely on the face, but at a distance, where the facial expressions are hardly perceived, the body becomes prevalent to understand and express emotions. For an overview of bodily manifestation of emotions, please refer to Kleinsmith and Bianchi-Berthouze (2013), Melzer et al. (2019), and Witkower and Tracy (2019). Unfortunately, there is no equivalent to a FACs for the body. An initial equivalent model, called Body Action Coding system (BACS), was proposed quite recently by Huis In 't Veld et al. in their two articles (2014a; 2014b). They investigated covert muscle activation across various body parts in the context of anger and

*Bridging the gap between emotion and joint action.* | Bieńkiewicz et al. (in press, NBR)

fear. Due to the lack of formal models from psychology and neuroscience fields, researchers in affective computing have hence turned to other fields for driving the design of automatic body expression recognition models. The four factors of the Laban Notation System (effort, shape, space, direction) (Laban and Ullmann, 1988) have provided the foundation for most of the pioneering work in this direction. A multi-layered approach inspired by Laban's Effort Theory for the automated recognition of emotion in dance performance was proposed by Camurri et al. (2003), through a computational model capturing how different dancers perform the same choreography with different emotions. Speed and energy showed to be correlated with the arousal dimension of the affective states while an extended body was generally associated with more positive states than a closed body posture. De Gelder and Poyo Solanas (2021, p. 1) define these features as middle level features and suggest that "*behaviorally relevant information from bodies and body expressions is coded at the levels of mid-level features in the brain*". A computational framework to model non-verbal emotions was proposed in the MEGA European project (Camurri et al., 2005) and a more recent approach was proposed in Camurri et al. (2016). By taking advantage of advanced machine learning architectures, there is today the tendency to use a more agnostic approach based on the temporal sequence of row data (e.g., Wang et al. (2021) from movement (e.g., 3D position of each joint, angles between body segments) or muscle activity sensors (e.g., intensity of muscle activity). Still, given the limited size of the datasets and also the complexity of body expressions when embodied in everyday activity, recognition performance may gain by a combination of low and middle level features. By being not so directly connected to the specific high-level semantic emotions, middle level features provide a more functional but still adaptive description of body expressions (de Gelder and Poyo Solanas, 2021) possibly enabling computational recognition systems to be able to generalize across different contextual situations.

Physiology. Physiological changes in relation to affect have been for long investigated. While pioneering work in the area of affective computing had initially leveraged medical devices, the technical advance in the low-cost wearable sensing technology has opened the possibility to seamlessly set up and explore ubiquitous applications (for a survey see Shu et al. (2018)). Differently from facial and body expressions, physiological changes are generally used to build systems that automatically infer affective changes along the valence and arousal dimensions. This is due to the lack of clear physiological patterns associated with discrete emotions (for a review see Siegel et al. (2018)). Applications for stress and anxiety levels automatic detection are possibly the most investigated areas in the computing domain (e.g., for a survey see Panicker and Gayathri (2019)). General approaches in building physiological-based affect recognition models built on general statistical features (e.g., max, mean, std) extracted from the physiological responses to an emotional event. To improve performances, more specific features are extracted for each type of physiological signals. Heart-related physiological activity is possibly the most explored in computing beyond galvanic skin conductance as it appears related to both valence and arousal. Features related to both the sympathetic and parasympathetic activity in both time and frequency domains have been explored (e.g., Alberdi et al. (2016)), for instance the ratio between high and low frequencies. While heart rate and skin conductance have been the most used physiological signals in the context of affective computing, respiration (Cho et al., 2019), skin temperature (e.g., Goulart et al. (2019); Wang et al. (2014)), and brain signals (Alarcão and Fonseca, 2019; Torres et al., 2020) have started to gain increasing attention demonstrating complementary performances. Similarly, research using electromyography (EMG) has shown evidence of muscle tension that is often linked to anxiety (Pluess et al., 2009). Indeed, in the work by Olugbade et al. (2019), the use of EMG in concomitance with motion capture leads to clear increase in automatic pain level recognition performances in people with chronic pain. This is again thanks to a technology that is more portable and acceptable for everyday use, enabling the extraction of continuous signals, extending the possibility for measuring a large set of statistical features and in particular features that characterize the variability of these signals (Cho et al., 2019). In a similar way to the work on non-verbal modalities, there is also the tendency to use advanced machine learning techniques that can work directly on raw data or on low-level statistical features extracted continuously over moving windows of the signals (Wang et al., 2021). However, this approach is still challenged by the limited size of the available datasets.

While each modality carries emotion information, studies have shown that multimodal recognition systems tend to lead to better performances (Al Osman and Falk, 2017; D'Mello and Kory, 2015; Poria et al., 2017). Since modalities work at different temporal scales as response to emotional triggers, how to fuse such modalities has been and is still a crucial question in the affective computing community. A variety of fusion approaches have been considered. Solutions have explored low, mid and high-level fusions of modalities, as well as more complex architectures to fully capture the relationship between such modalities. In particular, a typical issue in multimodal modeling is that some sensors may only be available during the training phases of the model. This could be due to sensor malfunctioning or sensor availability (e.g., privacy) during deployment. Some of the explored fusion approaches have tackled such problems by learning the relationship between modalities in order to infer the missing ones when the problem occurs (e.g., Cheng et al. (2016); Rivas et al. (2021); Wagner et al. (2011)). Transfer learning approaches have also been used to this purpose together with addressing the problem of limited dataset size (for a review see Feng and Chaspari (2020)).

*Bridging the gap between emotion and joint action.* | Bieńkiewicz et al. (in press, NBR)

However, these are not the only critical questions that challenge the affective computing community. Most of the work so far has focused on mapping face, body and physiological features or their combination into emotion semantic concepts. As we move into real-world applications, such approaches are quite limited as affective experience, and its perception, are subjective processes shaped by various factors such as context (Barrett et al., 2019a) and personality (Komulainen et al., 2014). Transfer learning approaches have been used to support the development of models between for example datasets built in the lab and smaller ecological dataset, or to compensate for the limited size of such datasets (Feng and Chaspari, 2020). Other approaches have more specifically attempted to integrate the context directly in the model. For example, the use of hierarchical architectures leveraging automatic human automatic activity recognition as contextual information to body expression recognition have shown to reach better recognition performances and generalization capabilities across a variety of activities (Wang et al., 2021). Such an approach was further supported by the use of graphical algorithms that intrinsically capture body configuration information critical to both the prediction of the activity performed and the emotion expressed by the body. Similarly, Zhao et al. (2019) have explored how personality can be leveraged to improve recognition performances of personalized emotion recognition models. Using a hypergraph learning framework, they captured the relationship between individual personalities and physiological responses to stimuli, showing a clear improvement in recognition rates, and suggested that the next step would be to co-learn the personality scores of participants.

While the work above is supported by the increasing number of multimodal and also multi-factors datasets, e.g., MAHNOB-HCI (Soleymani et al., 2012); DEAP (Koelstra et al., 2012); EMOPAIN (Aung et al., 2016), ASCERTAIN (Subramanian et al., 2018), there is the need for larger real-life datasets that are more inclusive and capture the variety of (social and activity) contexts, as well as the variety of emotional expression. While personality surely contributes to the experience, response and perception of emotions, there are many other personal factors (e.g., cognitive and physical impairment) that are critical to these processes. Existing datasets are still largely lacking the investigations of the above questions.

Multi-agent embodiment of emotion. One common characteristic of all the studies reviewed above, and of the-state-of-the art of embodied emotion in general, is that they all, almost exclusively, investigate the embodied manifestation of emotions in one individual in space and time (see Niedenthal, 2007). However, as said in our general introduction, humans are rarely withdrawn from natural interaction with other conspecifics. Particularly challenging is the issue put forward by leaders in emotion research (e.g., Ekman (1992)) that, arguably, the main job of emotions is to facilitate the engagement in perceived appropriate behaviors, in situational encounters with conspecifics and others. Dyadic and group situations are not only natural

*Bridging the gap between emotion and joint action.* | Bieńkiewicz et al. (in press, NBR)

vectors of emotion diffusion, they are the instances where this diffusion contributes to a successful communication and enhances prosocial behaviors. For instance, Mou et al. (2016) have shown that body behavior is a better predictor of emotion-group membership than facial expression, possibly because of the mirroring that may occur in group interactions. Playing music together is one of the most significant examples of non-verbal human interactive, creative and social activities, and, as music is widely regarded as the medium of emotional expression in full body movement *par excellence*, it is not surprising to witness the first layer of research focusing on emotion transmission in embodied joint action in this domain. For example, Glowinski et al. (2013) compared the expressive movement of violinists when playing solo and in ensemble conditions, and showed that when people perform a task as part of a joint action, their behavior is not the same as it would be if they were performing the same task alone. In the same vein, Varni et al. (2010) showed, in a multi-modal interactive context of a violin duo and a string quartet, that enacting pleasure while playing enhanced movement synchrony, whereas enacting anger reduced it. In another study with a triad of musicians, body sway was structured differently with different levels of emotional expressivity during performance (Chang et al., 2019). Higher Granger coupling within the triad of musicians (piano, cello and violin) was linked to emotional expressions of happiness when compared to sadness. Finally, the quality of dance performance was found to benefit from synchronized interpersonal movements, a quality that was also enjoyed by the spectators (Vicary et al., 2017). These examples illustrate the very recent move to understand human emotion in the context of joint action. However, the full picture remains obscure as we still have little to no understanding on how emotion dynamically fluctuates and propagates in multi-agent, naturalistic scenarios (where emotion brews as a consequence of interaction between agents and environment, e.g., Dotov et al. (2021)). Figure 2 represents summary points from the Sections 3A-3C showing how emotion is intertwined with acting together.

*Bridging the gap between emotion and joint action.* | Bieńkiewicz et al. (in press, NBR)

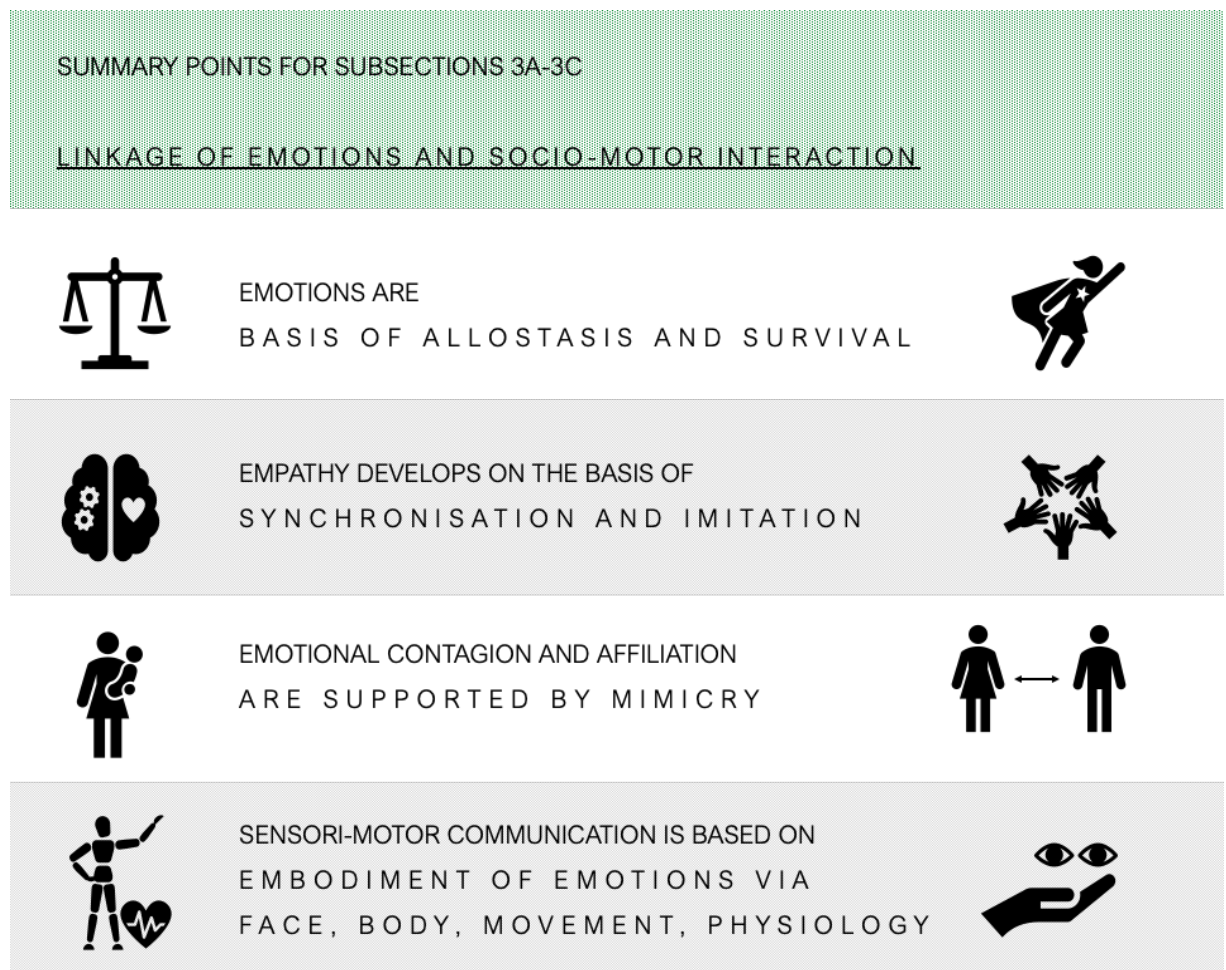

Fig. 2. Graphical summary of scientific findings reviewed in Subsections 3A-C and referred to in further sections of the manuscript.

## D. Linkage between joint action and emotion in socio-motor interaction deficits

Earlier (in Section 2D), we presented evidence showing how moving in synchrony with others can bring positive emotions such as affiliation and attractiveness. The impact of motor behavior and its shaping role for the emotions experienced by individual agents has also been brought to the spotlight by researchers interested in mental health and wellbeing (Macpherson et al., 2020). Strong incentive for further exploration in these domains comes from clinical research investigating psychiatric conditions - neurodevelopmental ASD and severe long-term disorder of schizophrenia (SCZ), which we now shortly review.

Clinical evidence from ASD studies: ASD is characterized by impaired development in terms of social interaction in general, communication and motor behaviors (American Psychiatric Association), with the underlying causes being still poorly understood. The ability to share attention with others, as well as imitate them, is pivotal for human development with the first signs of sharing experiences being recorded as early as in the first year of age in typical developing children (Kellerman et al., 2020; Tomasello, 2011). Those two adaptive mental functions enable symbolic play later in toddlerhood (Baron-Cohen and Cross, 1992), which lends itself to learning how to cooperate (e.g., turn-take) and communicate with others (Nadel, 2015). Hobson and Hobson (2007) proposed that ASD come from the difficulty to differentiate oneself from others ('theory of mind'). Fulceri et al. (2018) reported that ASD children synchronize with others better and imitate them more accurately (Jiménez et al., 2014) if the spatial goal for their own movement is clearly demarcated. This helps to draw a boundary with others. The evidence for the intact ability in ASD to imitate (Bird et al., 2007; Heyes and Catmur, 2021) is contradictory with studies reporting spectrum of difficulties with imitation and acting in synchrony (Baillin et al., 2020; Brezis et al., 2017; Fitzpatrick et al., 2017, 2016; Forbes et al., 2016; Koehne et al., 2016b; Marton-Alper et al., 2020; Tunçgenç et al., 2021; Williams et al., 2004). Impact of ASD on joint action during daily activities is also not clear from the scientific literature (Cerullo et al., 2021), with reports of children with ASD showing less interactional synchrony during naturalistic conversation with their partners (Zampella et al., 2020). Another study looking at the action of grasping a bottle, found that participants with ASD did not wait for their action partner and showed prolonged movements (Curioni et al., 2017). Trevisan et al. (2021) found that participants with ASD did not perform as well as their typically developing peers (measured as task performance and ability to sync steps with the interaction partner) in the collaborative task of carrying a table. In both reports (Curioni et al., 2017; Trevisan et al., 2021), difficulties with joint action performance were not linked to measures of other ASD-related motor deficits. In opposition to those reports, Scharoun and Bryden (2016) found no differences in healthy controls and ASD in joint dyadic tasks involving daily action (passing an empty glass of water to the researcher).

Complementary evidences exist in the brain neuroimaging literature. Studies using fMRI for action imitation and observation revealed that activity patterns for ASD in the MNS areas were altered compared to healthy control, along with networks involved in the social cognition and executive function (Chan and Han, 2020). Differences in neural activation patterns with healthy controls were also found with fNIRS during action observation and dyadic joint action in a block building task (Su et al., 2020a). Also, a recent facial electromyography study (Schulte-Rüther et al., 2017) demonstrated that although basic mirror mechanisms in ASD are preserved, they do not link to the high-order social cognition that allows emotion understanding and empathy. Difficulties reading facial expressions of negative emotions were reported for young males with ASD using EEG recordings (Van der Donck et al., 2020). Disrupted brain-to-brain coupling in ASD children was also found in the hyperscanning fNIRS study looking at joint dyadic action of ASD children with their parents during keypress tasks (Wang et al., 2020), but not in the study by Kruppa et al. (2021). Linked to this topic, we identified an fMRI study (Moriguchi et

*Bridging the gap between emotion and joint action.* | Bieńkiewicz et al. (in press, NBR)

al., 2009) looking at alexithymia in adults (self-awareness of emotions), demonstrating that people with alexithymia have higher activation in the MNS area, therefore showing similar difficulties of differentiation between self and others (i.e., a neural signature in the right superior parietal structure) to those found in the ASD population. Dunsmore et al. (2019) investigated the physiological linkage between interaction partners (heart inter-beat intervals) and found that the patients suffering from ASD do not sync their heart activity with a physical presence of another person in the room as observed in healthy controls (Scarpa et al., 2018).

In sum, a prolific state the art on ASD reveals differences at both neural and behavioral levels to neurotypical peers, in the ability to share attention during interaction with another person, to perform a joint task together, and to read and recognize their emotions. There is a contradictory evidence concerning the specific role of imitation ability in those deficits.

<u>Clinical evidence from SCZ studies</u>*:* Schizophrenia (SCZ) is usually diagnosed by the presence of negative symptoms, understood as social withdrawal and emotional flatness, and positive symptoms, understood as change in behavior or thoughts due to hallucinations or delusions. Green et al. (2015) reviewed the literature describing the difficulties with social interaction characteristics for SCZ and summarized them as deficits in empathy, reflective social processing - mentalizing, emotion regulation, face and voice perception. In parallel, the decrease in synchronization performances in people with SCZ compared to healthy controls was found in multiple studies, particularly for intentional synchronization (Manschreck, 1981; Varlet et al., 2012), predictive timing (Wilquin et al., 2018) and lack of sensitivity to social cues facilitating the synchronous performance with others (Cohen et al., 2017) or imitation (Sansonetti et al., 2020). In a pendulum-based synchronization task, Del-Monte et al. (2013) also showed that non-affected relatives of people with SCZ exhibit a decrease in synchronization ability, due to compromised visual tracking, pointing toward genetic SCZ phenotype for cognition. Interestingly, Raffard et al. (2018) found compromised stability of synchronization in patients with SCZ, but observed it to a lesser degree if the participants were positively primed to improve their sense of 'connectedness' to their task partners before performing the task. In a more ecologically set study by Kupper et al. (2015), participants with a paranoid type of SCZ revealed broken body synchrony patterns during dyadic role play conversation with healthy people. Synchrony negatively correlated to the social competence and severity of the negative symptoms, with participants not imitating the movements of their interlocutors, regardless of SCZ medication. Positive symptoms interacted with a lack of synchronous behavior by the healthy interlocutor, which could display affiliation, perhaps linked to the erratic movements caused by psychotic behavior.

*Bridging the gap between emotion and joint action.* | Bieńkiewicz et al. (in press, NBR)

Neurophysiological investigations of social and emotional syndromes in SCZ patients have also revealed interesting findings in the context of this review. For instance, searching for brain activation patterns in a fMRI scanner when observing recordings of finger movement and facial expressions, Horan et al. (2014) reported that people with SCZ did not differ from healthy controls, but showed a disconnection between brain activation and self-reported empathy reported through the Interpersonal Reactivity Index. In the same vein, Marosi et al. (2019) used EEG to investigate face and facial affect recognition in people with SCZ and revealed irregularities in the activity of the magnocellular pathway for face and face emotion processing. In a large study comparing MRI scans, Schilbach et al. (2016) reported differences between people with SCZ and healthy controls with regards to the connectivity in the MNS network and mentalizing network (left dorsomedial prefrontal cortex, left praecuneus, right and left temporo-parietal junction). Irregular connectivity in this area sheds light on the interpersonal difficulties that the patients with SCZ experience and on their ability to act with others. Together, these clinical studies, along with Subsections 2C and 3B, mount evidence that the MNS network might be a double agent for emotion and acting together.

*Bridging the gap between emotion and joint action.* | Bieńkiewicz et al. (in press, NBR)

SUMMARY POINTS FOR SUBSECTION 3D

IN ASD AND SCZ DIFFICULTY IN
EMOTION/INTENTION RECOGNITION AND
ACTING TOGETHER ARE INTERTWINED

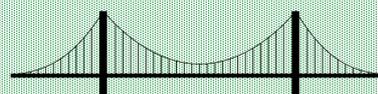

BOTH DIFFICULTIES ARE LINKED TO ATYPICAL NEURAL ACTIVITY IN
NETWORKS COMMONLY REFERRED TO AS
'MIRROR NEURON SYSTEM'

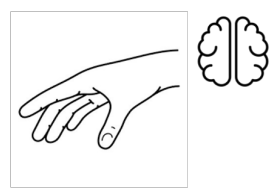

IN ASD DEFICITS IN INTERACTION ARE LINKED TO POOR ABILITY TO
SHARE ATTENTION AND DIFFERENTIATIE
SELF FROM OTHERS

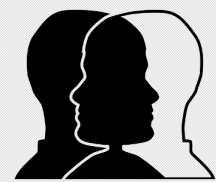

MOST EFFICIENT THERAPY AVENUES IN ASD/SCZ ARE BASED ON
SYNCHRONIZATION AND IMITATION TRAINING

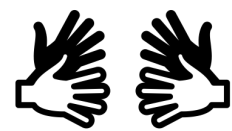

Fig. 3. Graphical summary of Subsection 3D highlights. Images: Suspension Bridge by Pechristener; Brain, idea, mind by iconfinder.com; Grabbing Hand by Oleksandr Panasovskyi/ Psychologist by Dirk-Pieter van Walsum from the Noun Project.

Interventions for ASD and SCZ: There is a great interest in using imitation as leverage for intervention studies for both ASD and SCZ. In an early intervention report, ASD children who received structured intervention focused on imitation and joint attention improved their social interaction skills, such as gaze following and requesting (Warreyn and Roeyers, 2014). Similarly, Landa et al. (2011) ran a randomized control study looking at the intervention targeted at imitation versus other therapy approaches long term improvement of positive affect and joint attention. (Koehne et al., 2016a) reported benefits of intervention for adults with ASD using a dance therapy program focused on movement imitation and synchronization which over three months improved emotion interference along with other abilities to imitate and synchronize with others. For individuals with SCZ who participated in the therapy sessions, involving imitation of others and other theory of mind components (with a control group in a

*Bridging the gap between emotion and joint action.* | Bieńkiewicz et al. (in press, NBR)

therapy focused on the problem solving skills) improved emotion recognition from the social situation and from the understanding of their intention of the movement (Mazza et al., 2010).

In sum, research in ASD and SCZ (both socio-motor interaction deficits) shows complex relationship between multi-faceted difficulties (such as differentiation from others, poor synchrony and imitation) and the ability to act together with others and understand their emotions (see Figure 4). Neuroimaging studies pinpoint differences in information processing in those populations related to the MNS network along with its linkages to higher social cognition and mentalizing networks.

## 4. Emoting joint action with non-human agents

Consistently over the last years, research in Human-Robot Interaction (HRI) and in Human-Computer Interaction (HCI) focused its efforts on developing social artificial agents that can initiate and maintain efficient and pleasurable interaction with a human. Emotional convergence benefits coordination between partners (Butler, 2015), and design studies in HCI have marshaled evidence that synchrony in movement qualities for facial and body expressions is essential for fluent human communication with virtual agents (Castellano et al., 2010; Marin et al., 2009; Numata et al., 2020).

Current developments in HRI are targeting real-time sensitivity to human expressions and behavior to promote long-term relationships (Castellano et al., 2008; Terada and Takeuchi, 2017). One of the main challenges is the capacity of the agent to operate on fast timescales, under one second, to be able to capture 'social moments' (Durantin et al., 2017). For instance, the *PEPPER* robot can infer possible interactive scenarios with customers via algorithms analyzing facial movement and voice signals in that time frame (Aaltonen et al., 2017). In other studies, robots adjust the interpersonal distance as a function of the estimated level of experienced emotions of the human in front of them (Bajones et al., 2017). The affect control theory offers a guiding principle used to create AI systems which are sensitive to affective states, adjusting their operations as a function of the context and need of their human interacting partners (Hoey et al., 2016). According to this theory, humans engage in situations that evoke emotions and feelings corresponding to one's culturally built affective span. In general, captured data are used to infer human affective states to which social robots adapt in various semi-autonomous ways.

Scholars in HRI have adapted and simplified the human emotional repertoire to social robots. For instance, the *ASIMO*, *JUSTIN* and *NAO* robots are programmed to express six basic emotions (and their various combinations): anger, disgust, fear, happiness, sadness, and

*Bridging the gap between emotion and joint action.* | Bieńkiewicz et al. (in press, NBR)

surprise. In general, human participants correctly recognize all basic emotions from the upper-body movements with a success rate of 75%-100%, with some exceptions however (van de Perre et al., 2015). Other robots, such as the *iCUB* robot (Metta et al., 2008), crawl or semi-autonomously manipulate objects in various dyadic contexts, and learn by doing and imitating (Billard and Dautenhahn, 1998; Boucenna et al., 2014).

Core research activities in the field of affective interaction with artificial agents have been established around two main populations sensitive to affective interaction: people suffering from social disorders (with a particular focus on children with ASD) and elderly people. Here we briefly summarize the state-of-the-art in these two domains, and then present recent trends in multi-agent and collaborative robotics.

## A. Social HRI and HCI in ASD research

As it is unrealistic to expect children with ASD to continuously and smoothly interact with affectively embodied robots, adjustments have been made to simplify the child-robot interaction and discriminate between positive and negative emotions, particularly to launch social interaction (e.g., Feil-Seifer and Mataric, 2011). The *PROBO* robot, for instance, imitates animal movements which helps ASD children to recognize basic emotions (Pop et al., 2013). While some robots are only able to detect emotion displayed by humans, others, such as *QRIO*, also depict facial expressions and include corresponding body manifestations of some emotions, for instance happiness and fear, in a way that is recognizable by humans (Tanaka et al., 2004). Another example of a robot expressing emotions is *MONARCH* (Sequeira and Ferreira, 2016). This is a companion robot deployed in children's hospital facilities and successfully integrated into a rich and complex clinical environment. In the related field of HCI, social robots are often replaced by virtual agents designed to create a specific social relationship with their human counterparts. Virtual and augmented realities are commonly used to help ASD children to focus on and recognize facial nonverbal cues (Chen et al., 2016), to learn to recognize and express emotions with their full-body movement (Alborno et al., 2016), to learn the required social skills (Lorenzo et al., 2019) or to promote verbal and nonverbal communication skills via joint actions (Srinivasan et al., 2016, 2015).

## B. Social robotics for the elderly

The ageing population is another major category targeted by research on HRI, at home (e.g., Fischinger et al. (2016)) or in nursing institutions (Moyle et al., 2013). Older people develop a wide variety of age-related conditions that can cause their vulnerability to minor stressor events and lead to loss of autonomy: this phenomenon is commonly known as "frailty". A number of interventions have been developed to target negative symptoms such as loneliness, anxiety and depression, which can also accompany dementia (Cifuentes et al., 2020; Kachouie et al., 2014; Valentí Soler et al., 2015). One example is a NAO robot-based rehabilitation program for people with dementia based in a geriatric ward, which reported higher outcome scores than conventional therapy on the immediate well-being and satisfaction (Rouaix et al., 2017). Similarly, *PARO* is a robotic seal that elderly residents in nursing homes benefit from by verbally interacting with it (Moyle et al., 2013). *HOBBIT* is another emotional assistive caregiving robot used at home to prevent falling (Fischinger et al., 2016). Further, a Social Assistive Robot exercise system was reported as more engaging the elderly in aerobic physical activity than virtual coach (Fasola and Mataric, 2012). The robot in Zhang et al. (2019) computes continuously the person's movement trajectory, while assisting with their dressing, but is not emotion aware. These are only five examples that have been extracted from a plethora of research and proof-of-concept studies and have demonstrated how useful HRI and HCI approaches can be in the clinical context, as well as at home, to accompany healthy aging.

## C. Environments of multiple human and artificial agents

When it comes to the environment being social, i.e., acting together in a group, a modest number of studies overcome the limitation of a robot(s) of working with more than one human, and only very few robots are adapted to such interactions. For example, the interactive robot *KEEPON* can engage in both dyadic and triadic interaction due to emotional expressivity which aids to build joint attention with the interaction partner, e.g., looking in the same direction as the human (Kozima et al., 2005). Besides *NAO,* which is known to be able to work as a guide in a museum for a group of visitors (Gehle et al., 2014) or with school-aged children (Hood et al., 2015; Ros et al., 2014), and *TIRO* which serves as a teaching assistant in musical classes (Han et al., 2009), the literature on human-robot group interaction remains scarce and almost exclusively in the form of "one-to-many" (a *star* graph as in the guide situation) in contrast to a more generic form of "many-to-many" (a *complete* graph, see Bardy et al. (2020)). One of the unique endeavors employed triads of BEAM robots in a semi-autonomous control mode (Wizard of Oz) during game playing scenarios with human triads (Fraune et al., 2019). Human participants reported changes in subjective fear and motivation moderated by the perceived cohesion of the robot group, in comparison to other typologies with one human versus three robots and vice versa, and one-to-one interaction between a human and a robot. This indicates a breadth of emotional component to be explored in intergroup dynamics between human and artificial multiple agents, despite robots not being embodied with sensori-motor communication abilities (embodiment of emotion). However, despite being 'emotionally neutral', in a study by Kochigami et al. (2018) robots NAO and PEPPER played social roles by creating social ties between human group members (children and adults) and successfully facilitating interaction between them. Examples of similar studies are limited in number (see Figure 4). Sebo et al. (2020) pinpointed key messages emerging from the current state-of-the-art: (i) behavior in one person to one robot does not interpolate on the group behavior; (ii) verbal and non-verbal robot behavior shapes the response within the group and can support cohesion; (iii) people are more likely to engage with a robot when they are in groups; (iv) similarity (anthropomorphism) to humans plays a role in the integration of robots in a group. The dynamics of how emotion can be shared or propagated through the heterogeneous networks of several humans interacting with several artificial agents is unknown and has profound implications for the future of collaborative robotics.

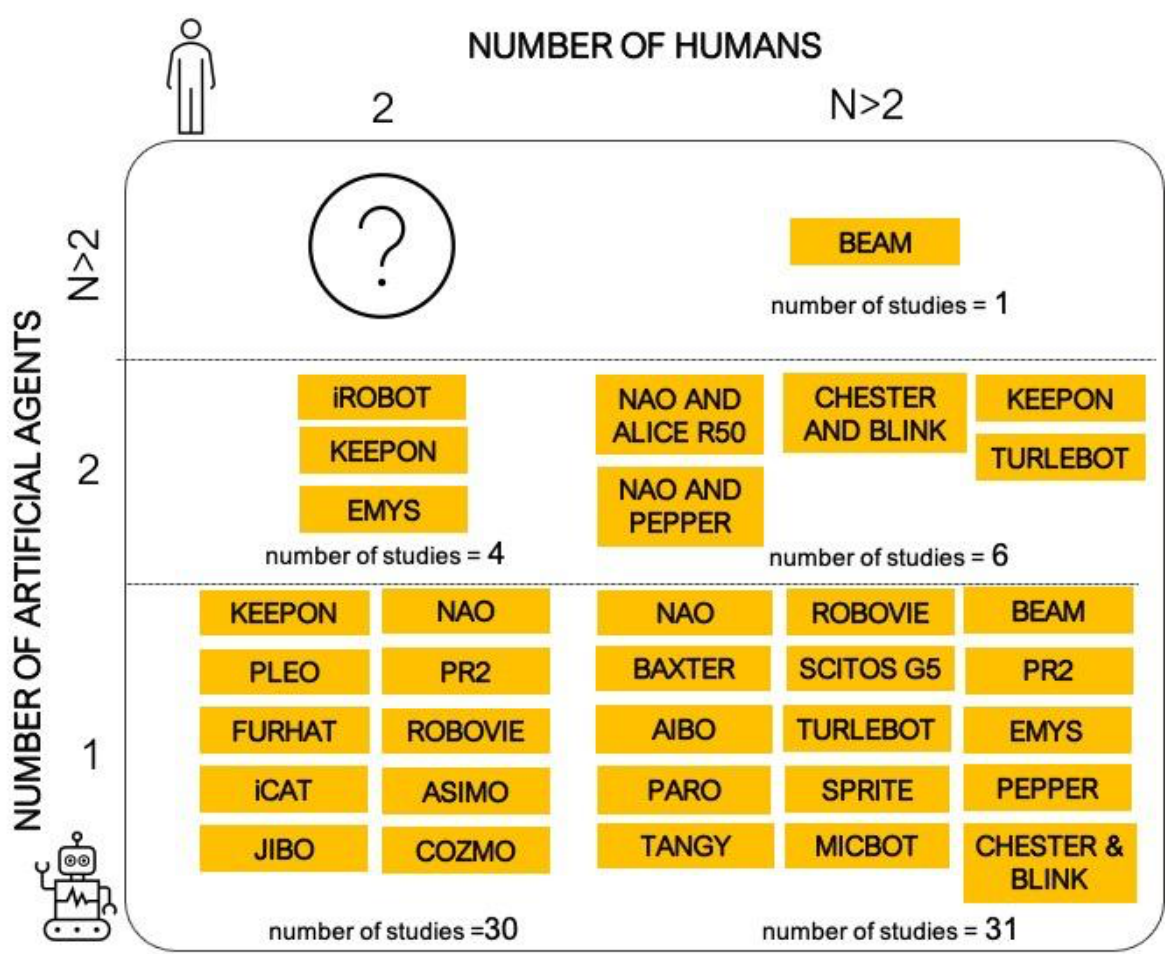

Fig. 4. The 2D matrix of the number of agents (persons ≥ 2 and robots) in HRI interaction extracted from the literature review of Sebo et al. (2020). Examples of the names of the robots used are placed on orange tiles with the count of overall studies identified by the researchers (refer to Table 1 in Sebo et al. (2020) for further details).

## D. Collaborative robotics in the industry

Emotion sharing as a means to facilitate social interactions in HRI has so far mostly been applied in therapeutic settings (see Subsections 4A-B). In industrial settings, however, despite the heavy reliance on industrial robots for manufacturing, such examples are still rare. It is believed though that an essential component of the next industrial revolution, often referred to as Industry 5.0, will be that of the collaborative robot, a robot that can complement human co-workers, performing tasks that are either tedious or dangerous (Demir et al., 2019). The reason is because industrial sectors still lacking in terms of automation are those that cannot be fully automated, as they require human participation (Elprama et al., 2016). It has been recognized, however, that introducing collaborative robots to workplaces might have an adverse effect on social interactions in these workplaces. Untrained personnel, in particular, tend to expect the same social signals from robots as they would from human colleagues and expect the robots to adhere to existing social practices (Fischer, 2019). If collaborative robots fail to understand social signals and respond accordingly, they will be seen as impolite, cold and uncooperative. It also represents a missed opportunity to convey the robot's capabilities, while making communication more dependent on disruptive explicit signals, when more fluent implicit signals would have been preferable for seamless collaboration (Breazeal et al., 2005). Fischer (2019) further argues that collaborative robots do not only need to understand and produce social signals but that these signals need to include emotional expression. The reason, as seen in the above sections, is because sensori-motor communication of emotion and intention is an integral part of conventional social practice. A robot is simply expected to be sad when delivering bad news or happy when successfully completing a challenging task. Emotional expression may also be used to communicate real needs, such as when the system is running out of power and needs to be recharged. Recognition of human emotional expression under natural industrial conditions is difficult, as the technology needs to be both non-intrusive and robust over time. Speech (Khalil et al., 2019), gaze (Admoni and Scassellati, 2017) and facial expressions (Li and Deng, 2020) become more convenient cues than gestures or full body movement (Liu and Wang, 2018). However, with the introduction of cheaper wearable sensors, emotion recognition from EEG has recently become a viable alternative (Toichoa Eyam, 2019; Zheng et al., 2019). For expression of emotions, collaborative robots are limited by their embodiment and interfaces have more often been used for conveying information than for social signaling. Most collaborative robots only rely on projections of faces on flat screens to express emotions (Kalegina et al., 2018), if such expressions are used at all. There are recent examples, however, where the embodiment has in fact been exploited for social signaling, even highlighting the importance of a breathing motion (Maric et al., 2020; Terzioglu et al., 2020), opening new emotion-based perspectives in collaborative robotics.

# 5. Avenues for future research

Sections 1-4 presented ample evidence for the interplay between emotions and joint action. Humans are unequivocally attuned to each other, with body movement being a powerful carrier of idiosyncratic information (Cutting and Kozlowski, 1977; Loula et al., 2005; Troje, 2002) and socially meaningful qualities (e.g., Centelles et al., 2011; Clarke et al., 2005; de Gelder and Poyo Solanas, 2021; Nackaerts et al., 2012). Information about the arousal of a person encoded in movement and their intention can be transferred to another person, for example, as a forewarning of a threat. Being able to read out those non-verbal signals from others, along with the ability of humans to couple their body and brain activity, is the cornerstone of successful communication and cooperation between people. We strongly believe that interaction is key to understanding the human brain, as the human brain, through interaction with the environment, is of a physical, but also fundamentally of a social nature (Section 4 recaps why this also applies to hybrid interactions between humans and artificial agents). We acknowledge, as do other researchers, that social interaction should be at the forefront of neuroscience research (Schilbach et al., 2013). Some recent attempts, such as the social alignment theory, using herding modeling by Shamay-Tsoory et al. (2019), are providing other important milestones towards this venture.

In this last Section (5) of our review, we put forward the idea that emotional arousal should be considered as an integral part of the so-called 'motor system', shaping and fine-tuning the real-time socio-motor interaction with others. Emotions arise as responses to the stimuli in the environment (with a function to maintain/restore allostasis) and bear subsequent impact on one's perception, affective state, ongoing and future movements (e.g., Wood et al. (2016)). Thus, we propose to incorporate emotion in a joint action context as one entity, a 'third eye' that steers other mental and physiological processes to navigate the rich, multi-modal layers of the multi-agent social space. As emphasized before, the scientific evidence on the emotional embodiment during socio-motor interaction is limited (see Section 3C for overview), especially in terms of studies exploring real, not enacted, emotional arousal with naturalistic scenarios as a backdrop. To disentangle and decode the emotion propagation and socio-motor interaction not only via movement, but also via physiological processes, we propose to follow new research avenues (Subsections 5A-E) to decipher the unknowns about interplay of emotion and joint action (see below the research questions we have identified as a part of literature review process, continued further on Fig. 6.).

Research questions: We now know that some features of body posture and movement carry emotional qualities (de Gelder and Poyo Solanas, 2021), but can we find a motor signature of emotional arousal and valence in body movements in the context of socio-motor interaction (regardless of particular body parts) specific to i.e., particular levels of valence or arousal? To what extent context and culture shapes this signature during joint action? Is there a group emotional signature emerging from individuals sharing space at the same time (i.e., euphoria of football fans in the tribunes) (Subsection 5A)? How do emotions evolve and propagate through the network of agents (humans or hybrid groups of human and artificial agents) and if so, how do they influence the outcome of joint performance (Subsection 5B)? Further, we dive into the need for adoption of multiple timescales approaches, which emerged throughout this review (Subsections 2D, 3B-D, 4), showing that environmental function of emotion unfolds over multiple time windows, throughout physiological and movement qualities, as well as that some factors in socio-motor interaction are only meaningful when looked through an appropriate temporal lens (i.e., expertise, culture, previous experience with an agent). What are then the crucial timescales we need to integrate into the research agenda to advance this enquiry (Subsection 5C)? How can AI techniques assist in this process, and unravel the patterns of information about emotion/intentional qualities and its propagation in agents during joint action (Subsection 5D)? Finally, we dive into the world of digital, currently disembodied interaction, and highlight the notions that recent pandemic experience has left in terms of interplay of physical presence and emotion embodiment during social interactions (Subsection 5E).

## A. The notion of Emotional Motor Signature in joint action

A large body of the work reviewed in Section 3C cements the foundations of the embodied nature of emotions, according to which expressions of emotional dimensions, rooted into common neurophysiological structures between cognition and action, are diffusing into movement physiology and are visible in facial, distal, as well as proximal parts of the body (Barrett et al., 2019b; de Gelder and Poyo Solanas, 2021; Kleinsmith and Bianchi-Berthouze, 2013; Melzer et al., 2019; Witkower and Tracy, 2019). Interestingly, this extensive literature does not yet intersect with another parallel body of research developing concepts and methods to assess Individual Motor Signatures (IMS), the idiosyncratic way each individual moves (e.g., Słowiński et al. (2016)). Pioneering work of Johansson (1973) put forward the notion of transformational invariants (Malcolm, 1953), i.e., that persistence in some dimensions (e.g., length, ratios) across the motion of others (e.g., global transformation of the local optic flow) could help observers to quickly extract person-related properties, very relevant to socio-motor interaction context. IMS often relies on movement velocity as a key feature, as it is both stable across time and repetitions for each individual (movement similarity) and differential between individuals (inter-individual movement difference). Differences in the way people move during the performance of a motor task can be captured by using 95% confidence interval ellipses in the similarity space (Słowiński et al., 2016). This is an abstract two-dimensional geometrical space minimizing distances between repetitions and individuals by using ad-hoc dimensional reduction techniques. Ellipses can be large or small depending on intra-individual variability and can be close or distant from each other depending on between-individual variability. The approach has proven useful in identifying IMS in various populations, ranging from healthy individuals to people suffering from schizophrenia (e.g., Słowiński et al. (2017)). It has also proven useful in various tasks and contexts such as in the mirror game (Słowiński et al., 2016) or during improvisation movement (e.g., Coste et al. (2019)), and at different distal or proximal (and more postural) parts of the body (e.g., Coste et al. (2020)). Whether IMS, {as well as Group Motion Signatures (GMS, inter-group movement differences), i.e., the way IMS are assembled together in an ensemble of individuals engaged in reaching a common goal during joint action}, are emotionally neutral, and whether they are of different shapes and locations in the similarity space when produced in various emotional contexts (intra-subject variability) remains open to investigation (see Lozano-Goupil et al. (2021), for a first evaluation of Emotional IMS). Taking this road would not only answer the above questions, but would also offer a way to reconcile existing theories of emotion and those of embodied social interaction, inclusive of intra- and inter-individual/group variability and concepts such as motor accents (e.g., Ting et al. (2015)), in a real-life context of a joint action.

## B. Emotional group synchronization models

As emphasized in Section 2B, models of perceptuo-motor social synchronization when $N>2$ have not yet incorporated emotional qualities in their constituents, i.e., they remain emotionally neutral, despite the evidence gathered in this review that emotions are contagious, propagate through the social network, and constitute the essence of joint action. One urgent avenue of research requires a complementary approach incorporating emotional qualities in experimental and modeling scenarios. On the experimental side, the manipulation of positive, negative, and mixed emotional qualities, be they enacted or (ideally) induced, and the observation of how these qualities propagate from one node to the next across the collective sensori-motor network, converge or conflict, is requested. On the modeling side, coloring coupled oscillator models of synchronization with those emotional qualities would help to better understand, and generalize, the underlying propagation mechanisms. For instance, the network of coupled Kuramoto oscillators presented in Section 2B, capturing group synchronization regimes when perception is present (see Bardy et al. (2020) for details), needs to be adapted to incorporate emotional qualities at individual levels, such as:

*Bridging the gap between emotion and joint action.* | Bieńkiewicz et al. (in press, NBR)

$$\dot{\theta}_i(t) = \omega_i + c \sum_{i=1}^{N} a_{ij} \sin \left( \theta_{jEM}(t) - \theta_{iEM}(t) \right)$$

where *N* is the number of agents, $\theta_{iEM}$ the phase of the movement of the *i*-th agent under emotion *EM*, $\omega_i$ their natural frequency, and *c* the strength of the coupling with the other agents when perceptual coupling is established. Coefficients $a_{ij}$ are set to a value between 0 and 1, depending on the dyadic perceptual coupling between agents *i* and *j*, i.e., the spatial configuration of the group. Coloring such oscillatory models with emotion-aware individual signatures (see Figure 5, Example C), nourished by experimental data, would therefore be an operational way to close the gap between Sections 2-3 of the present review.

## C. Embodied emotion across multiple timescales

In this review, we have hinted at the concept of multiple timescales on a handful of occasions. In Greek pre-Socratic philosophy time was represented by two notions; Chronos, sequential and linear time as we currently understand and apply it in a metric system (chronological time), and Kairos; which resembled the 'right' time, especially in the context of an action affordance (i.e., time for harvest). A myriad of research studies has provided data-driven rationale in favor of the use of multiple timescales to capture animal behavior and physiological processes (be it signal duration, temporal resolution, units applied or temporal dynamics). The evidence for multi-timescale behavioral organization has been recently investigated in *C. elegans*; showing how neural dynamics in this much simpler organism (slow - low frequency; fast - high frequency) orchestrates different movements and allows for flexibility of behavior (Kaplan et al., 2019). In humans, communication is regarded as a robust example of multi-modal behavior stretching across multiple levels of temporal structures due to the variety of interconnections between internal systems such as respiration and movement (Bardy et al., 2015; Pouw et al., 2021).

<u>What do we know about temporal aspects of emotions?</u> Since the works of Solomon and Corbit (1974) it has been widely accepted that emotions unfold their dynamic over time, rather than spike events, with complex temporal structure (De Gelder et al., 2004; Frijda, 2007). Regardless if the stimulus is aversive or hedonic, the response curve for high arousal physiological reactions (i.e., heartbeat) unfurls as (i) rise to peak, (ii) adaptation period, (iii) recovery with reversed peak to re-establish baseline within 30-60s. In a LORETA EEG paradigm, Costa et al. (2014) found a precise pattern of neural signatures of fear, disgust, happiness and sadness, with differences emerging mostly in the temporal characteristics of neural activation, but not the spatial spread. The differences that emerged are as follows: (i) Early onset (around 200ms post exposure) and shorter duration characterized emotions - fear and disgust, which are associated with a need for quick body reaction; – (ii) Early onset (around 260ms post stimuli) with a second processing peak at around 400ms in different areas - happiness; and (iii) Late onset (around 400ms post exposure) and longest duration (90ms) – sadness. Personal diary study (Verduyn et al., 2015) reported similar temporal patterns linking to the adaptive behavior evoked by emotion, meaning that fear and disgust operate on fast timescales as they require quick fight-or-flight reaction; whereas emotions like anger or joy, on average, take longer to disperse through action (Costa and Crini, 2011; De Gelder et al., 2004; Feldman Barrett and Finlay, 2018; Frijda, 2007). Notably, sadness was earmarked as the longest in duration, perhaps because its presumed function is to pave a pathway to rumination, motivation for change in personal circumstances or acceptance. Personal dispositions, from reactiveness and resilience on a physiological level to higher cognitive functions such as emotion regulation processes (reappraisal), were highlighted as subjected to inter-subject differences (Solomon and Corbit, 1974)

How emotions influence motor timing? Distal movements in particular (i.e., object manipulation) are a clear *reservoir* of emotions, a cornerstone assumption of forensic criminology. Gao et al. (2012) analyzed movement in a touch-based game and found (i) the length of the stroke to be indicative of the dimensional quality of valence, (ii) speed and direction to be indicative of arousal, while (iii) pressure specifically discriminated anger from other states (where increase in energy transmitted to movement has functional significance). Similarly, more frequent manipulation of the computer mouse was found to be associated with higher stress (Hernandez et al., 2014). During paced synchronization, adults and children tapped faster if they were primed with negatively valenced pictorial stimuli before the trial (Monier and Droit-Volet, 2018). The speeding up of motor response was interpreted as activation of the fear circuitry evoked by negative emotional induction (LeDoux, 2014), leading to the speeding up of the internal clock system (Cheng et al., 2016) and shifting movement towards faster timescales. Further, emotional arousal leads to subjective perception of time in some tasks (Gil and Droit-Volet, 2012), making a point that time perception is tied to the quality of stimuli (Grisey, 1987). This becomes particularly relevant in the context of untangling the dynamics of joint action and emotion (i.e., during synchronization).

What do we know about group temporal dynamics? Vesper et al. (2011) has demonstrated that better coordination is achieved in dyadic action when participants make themselves more predictable (less temporally variable), in comparison to performing identical (pointing) movement alone or next to another person without intention to act together. Grammer et al., (1998) demonstrated that opposite sex pairs show a complex temporal structure of interaction patterns of body movement during conversation, which is repeated if both sides show interest, and is unique for each dyad. In a previously mentioned model of a psychotherapeutic alliance (Koole and Tschacher, 2016) three timescales were proposed for interpersonal synchrony, namely; (i) a *phasic* time-scale, which runs from a few hundred milliseconds to 10s, characteristic of motor synchrony, (ii) a *tonic* time-scale, which runs from about 10s to an hour, and involves more complex forms of social cognition, such as language and reasoning, and (iii) a *chronic* time-scale, stretching from weeks to years, and which involves the development of complex emotion-regulatory abilities. Bardy et al. (2020) reported that social memory (expertise in dance practice, related to (iii)) can affect the ability to synchronize with others under different perceptual strains. Similarly, experts in capoeira and tango had higher ability (kinesthetic ability) to imitate and synchronize with others, in comparison to athletes who also practice group sports, but without the synchronization component (Koehne et al., 2016c). More broadly, Burgoon et al. (1995) suggested that behavioral norms can pass from one generation to another as culture (i.e., think of the jovial behavior expected from a salesman versus the stoicism of a medical professional). An unexplored territory is investigation of previous personal experience with agents partaking in socio-motor interaction, which can trigger certain emotions prior or during the execution, due to predictions of the internal model (Barrett, 2017b).

Taken together, these findings hint at a hidden hierarchy of socio-motor interaction, from low-level, fast timescales, which are more appropriate for immediate behavioral synchronization and responses, to high-level, slower timescales, which involve complex cognition and emotion regulation, linked to perception of emotional qualities and social memory. In this context, future research should apply a multi- (or inter-) modal multiple timescale approach for studies set on the interface of emotion and human joint action.

## D. Leveraging artificial intelligence (AI) methods to capture emotions in socio-motor interactions

AI offers powerful analytical tools to deal with complex data, and so it is a valuable method for investigating individual and group motion signatures of emotions within the context of joint action. As discussed in Section 5C, a multiple timescale approach is required to build a solid foundation of how various emotional qualities propagate in joint action could provide a window into relative aspects of time and how it is linked to the differentiation of emotional qualities in movement and the opportunity to act collectively. However, AI methods that address multiple timescales are largely limited to encoding each temporal variable (e.g., each different modality or modality dimension) separately. Such methods can only really create models tuned to a single timescale per temporal variable. Perhaps the more promising direction is methods that capture multiple timescales within each variable itself.

The few studies in this direction (e.g., Gurcan and Nguyen, 2019; Ma et al., 2019; Yamashita and Tani, 2008; Yang et al., 2020) have so far been constrained to individual action modeling. Two of the few exceptions are the multiple timescale recurrent neural network (MTRNN) (Yamashita and Tani, 2008) and the Approach Group Graph Convolutional Neural Network (AG-GCN) (Yang et al., 2020). Whereas the AG-GCN was designed to model group behavior, the MTRNN was originally developed for individual action scenario but has been extended to dyadic (Hinoshita et al., 2009a) and group interaction (Jaderberg et al., 2019), although only for robots in simple robot-robot interactions and bots in a multiplayer computer game. There is not much analysis of the behavior of the MTRNN in the multi-agent settings, but a functional hierarchical structuring of events in the movement sequences sampled was shown to emerge through modeling at multiple timescales in its original use (Yamashita and Tani, 2008). Similar findings reported for multiple timescale AI architectures explored in the context of natural language processing (Chung et al., 2017) underline the value in pursuing multiple timescale modeling for further understanding emotion in joint action. Moving forward, there is the need for new architectures that simultaneously: 1) have pathways for both action and emotion recognition, drawing from neuroscientific findings of multitask coding of observed action (behavior identification and semantic interpretation) in humans (Gallese, 2007; Iacoboni et al., 2005); 2) with multiple timescales processing for the two, rather than for one or the other (such as in Hinoshita et al. (2009b)); and 3) capture individual as well as group signatures in group settings. Such architectures have the potential to maximize the value of AI for different aspects of emotion expression, perception, and propagation in the context of joint action.

In the current age of deep learning where low-level layers are entrusted with the extraction of signatures (i.e., features) from continuous streams of sensor data, there is a shift away from focus on hand crafting of computational features, with increasing significance instead placed on the design of architectures that learn relevant features organically and directly from sensor data. A more valuable approach may be a blend of both methods. On one hand, automatic feature extraction sub-architectures could perhaps lead to (deeper) insight into what behaviors or other responses at individual and group levels characterize emotions in joint action. Contemporary understanding of these individual and group signatures could, on the other hand, be further explored by employing them in the form of hand-crafted features. The large number of existing studies on affect recognition (see Section 3C for a discussion of some of these) would be valuable in guiding the choice of individual level features to examine in the context of joint action. The minimal set of studies (Mou et al., 2015; Ukita et al., 2016; Yang et al., 2020; Yücel et al., 2013), that have used group relations features for affect recognition and

*Bridging the gap between emotion and joint action.* | Bieńkiewicz et al. (in press, NBR)

related AI areas highlight differences in distance, speed, and direction (as well as displacement and/or velocity) between the individuals in a group as additional features to consider. As anticipated by some studies (e.g., Ukita et al., (2016); Yücel et al. (2013)) the problem of determining (the extent of) a group of interest is a challenge that may need to first be addressed, especially for autonomous AI to be integrated into real world settings, before understanding how emotion experience in joint action can be possible (for an overview of the current state-of-the-art on emotion recognition on groups please see Veltmeijer et al. (2021)). Such an AI system would need to be able to determine if there is any joint action group present in a given location or sequence of events, the number of such groups, and the membership of each of them.

## E. Emoted disembodied joint action in the digital world

The COVID-19 pandemic has speeded up our move into digital encounters across all aspects of our social life, be it work, education, leisure or even health. *V*irtual interactions have enabled millions of people to continue working together remotely (video calls e.g., Zoom or virtual spaces with people represented by avatars e.g., *Sococo* or *Virbela*). However, these virtual interactions have been impoverished in emotional context due to the lack of information, coming from gestures, body postures and facial expressions, about emotional arousal and agent's intentions. These non-verbal cues are critical to communication, understanding and bonding, recently captured by de Gelder and Poyo Solanas (2021) as mid-level qualities. Musicians, teachers and athletes across the globe have experienced how different it feels to perform without an audience feeding back their reception. After all, part of our identity comes from socio-motor interaction with others, by which we can express our personal qualities, such as being funny or having a preference to lead or follow. Others reflect our qualities via sensorimotor communication, which is the foundation for validation and updating our sense of self. The lack of such easily accessible non-verbal cues in virtual spaces and the amplification of facial and gaze cues over body cues in video calls (e.g., *Zoom*) have been suggested to contribute to "*Zoom* fatigue" (Bailenson, 2021). In sum, lack of socio-motor interaction with others deprives our brain from the habitual process of predicting the unfolding of their actions in order to efficiently affiliate and cooperate with them in real time. Given that a digital environment will most probably last to a certain extent post COVID-19, it opens the opportunity, but also calls for embedding and facilitating joint emotional interaction to become effective. How to enhance the communication of emotional expressions in virtual spaces has been previously investigated. However, these studies have been limited to simply manually expressing such states (e.g., Pita and Pedro (2011) showing that in such situations people spend more time in carefully crafting verbal affective expressions than they do in gestural ones,

*Bridging the gap between emotion and joint action.* | Bieńkiewicz et al. (in press, NBR)

possibly because of the lack of embodiment of the latter. Sensing technology, affective computing and sensory interactions or substitution research can have a crucial role in creating and sharing a sense of agency, a felt embodied affective state and at the same time advancing our understanding of how emotions become joint experiences. Leithinger et al. (2014) have shown how our own hand gestures can be transferred to another physical space as 3D objects for the others to experience in action. Remote tactile interactions, through the use of wearable devices that stimulate the other person skin in response to a remote tactile gesture (e.g., tactile exchanges in Huisman et al. (2013), such as skin stretching or being pinched (Alhuda Hamdan et al., 2019; Muthukumarana et al., 2020) could help maintain the affective power of our non-verbal behavior during remote communication. Unfortunately, none of these studies have yet explored how such approaches are suitable to transfer the emotion qualities of an action and definitely not how such emotion qualities transfer across a group. Instead, transfer and group dynamics have been explored through disembodied representations of emotion-related signals (e.g., galvanic skin response or HRV) or inferred emotions through computational algorithms (Ardizzi et al., 2020; Gashi et al., 2019). Also, very little attention has been given to the spatial and temporal aspects which characterize joint emotional experiences, which are becoming even more critical than before. Studies have also shown that the perception of self-location can be altered through the right manipulation of sensory feedback, as in Lenggenhager et al. (2007). As Nadler (2020) highlights, space takes new meaning and creates new affordances in these virtual spaces that alter the meaning of joint interaction. From a computational perspective, modeling group emotion may require us to integrate in the computational models the dynamic characteristics of such virtual spaces that are affected by their properties and typology of information flow (Bardy et al., 2020).

## 6. Summary

Emotions move us across multiple levels of qualities and timescales, for our own survival and higher, collective purposes. The sheer physical presence of others in shared space and time fulfills the most primal of human needs, which is to belong to a group. Recent pandemic experience (COVID-19) has demonstrated a devastating effect of disruption of routine social interactions. Joint actions have been obstructed by social distancing measures or by being moved entirely to the digital world. Confinement has had profound and not yet fully understood effects on mental wellbeing across all age and gender groups, (Ammar et al., 2020), and has had impact on the development of social skills in children deprived of contact with their peers (Giménez-Dasí et al., 2020). In this review we have highlighted a need to close the gap in the research between emotion and socio-motor interaction across different disciplines, and have prompted specific questions to the scientific community to do so. Although various branches

*Bridging the gap between emotion and joint action.* | Bieńkiewicz et al. (in press, NBR)

of science have separately focused on joint action and on emotion, there is a growing necessity to understand how emotions flow across our embodied social interactions, and how they affect us as individuals, as a group and as a society.

Figure 5 depicts dimensions that were identified in this review as meaningful to obtain a full picture of emoted socio-motor interaction, inspired by the research "landscape" of second-person neuroscience, proposed by (Schilbach et al., 2013). Sections 2-3 are represented in the lateral panel as emotion depicted a shade gradient of agents engaged in interaction (where color denotes interaction type), and organized the possible consequences of socio-motor interaction, into three 'working' categories that emerged during our literature search: performance, social and individual.

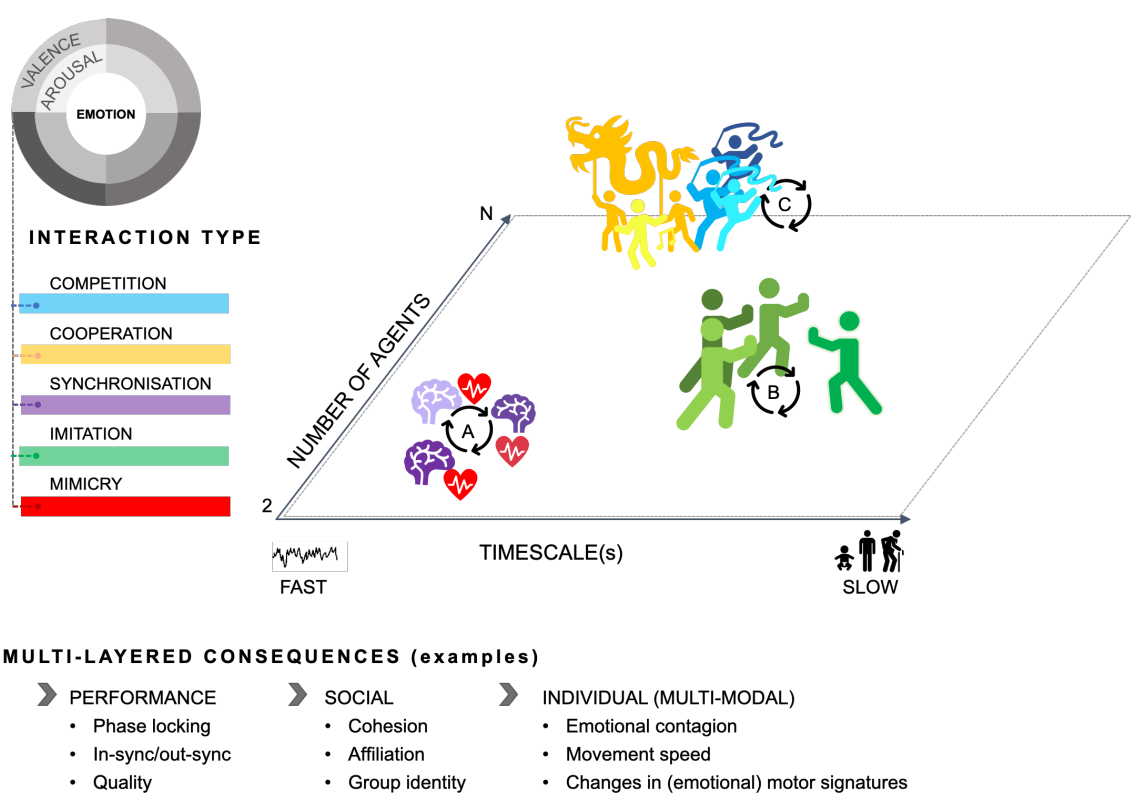

Fig. 5. Presentation of the future research landscape for emotion and joint action. **The bottom horizontal axis** represents **multiple timescales** that can be extracted from data (ranging from *ms*, i.e. brain and neural activity, to *hours and years,* i.e. expertise). **The vertical axis** denotes the number of agents engaged in socio-motor interaction, being perceptually and motorically active in the same physical or virtual space. **The colorful legend on the left side** represents possible **types of socio-motor interaction** that emerged from the literature review (delineating different spatio-temporal relationships between agents - see Table 1 for summary). **The circle in the left corner** represents models of emotions (gradients relate to dimensions of valence and arousal) that need to be adapted to multiagent scenarios, and here are injected into a color scheme of interaction types. **Bottom panel** lists **multi-layered consequences** of **the socio-motor interaction** across three main categories:

*Bridging the gap between emotion and joint action.* | Bieńkiewicz et al. (in press, NBR)

performance (quantitative and qualitative), social (i.e., affiliation and cohesion) and individual (impact on personal motor characteristics and emotional contagion from interaction with other agents), identified from this review. Example A – recordings of brain activity (hyper scanning) and heart activity during a naturalistic conversation between three people; Example B - A group is trying to follow and repeat the pattern from the leader (Tai Chi class), captured with motion and respiration recordings; Example C – Group of agents dancing in and out of synchrony with each other (music street festival/parade), intertwined with periods of coordinated competition with each other, captured with motion recordings, following different typologies (Bardy et al., 2020). *Images: The Tai Chi and Old Man icons come from* http://www.flaticon.com *(author: Freepik).*

Recent models of emotion (i.e., (Barrett, 2017b)) have paved a theoretical path to integrate aspects of the presence of others and acting together in order to bridge a new, more informed, interdisciplinary avenue of research that is inclusive of dynamic relationships between emotion and joint action performance, when more than one agent is present, and of action context. The scientific evidence gathered and synthesized in Sections 1-3 of this review provides a weighty incentive to embrace more holistic and interdisciplinary approaches that are built on the assumption that our brain is primarily predictive, over reactive, and that our emotions are based of interoception and exteroception, and play an important allostatic function. Human abilities to understand emotions and to act together develop simultaneously throughout the lifespan and show neural overlap in brain activity suggesting that both have been shaped by evolution to be interdependent. Therefore, the unraveling of the intrinsic relationship between emotion and socio-motor interaction needs to be built on modeling multi/inter-modal emotion propagation models, conceptualizing what group emotion is, and whether some emotions are exclusive to interaction (see Fig. 6).

OUTSTANDING QUESTIONS

- How do individual emotions shape the way we move during joint action?

- Does joint action require agents to tune into the emotional state of other group members via empathy or to acknowledge it via mimicry?

- How can we track multi(/inter)-modal propagation of emotional embodiment (i.e. qualities/features)?

- Are certain emotions amplified by being with others while other emotions are exclusive to being with others (i.e., timidness, shame or desire)?

Fig. 6. Questions we have not found answer to during our literature review and need to be addressed by inter-disciplinary research.

*Bridging the gap between emotion and joint action.* | Bieńkiewicz et al. (in press, NBR)

This will nourish new modes of social interaction with non-human agents, which can provide personalized care, entertainment and life-long education for fragile populations. We argue that deployment of machine learning algorithms and models supporting multiple timescales approach might provide the apposite caliber of research machinery to advance our comprehension of the dynamics of this 'dark matter' (Schilbach et al., 2013) of research and dissect the multi-layered physiological, socio-behavioral consequences of acting together, inclusive of the emotion component and cohesion between agents. Progress in this cross-disciplinary field will feed future research and development in many technological areas such as collaborative robotics for industry and healthcare (i.e., incorporating the design of artificial agent principles, sensors and effectors for social-signaling and sensori-motor communication), and will provide the tools for embodied, digital interactions (for virtual workspaces and education).

*Bridging the gap between emotion and joint action.* | Bieńkiewicz et al. (in press, NBR)

*Bridging the gap between emotion and joint action.* | Bieńkiewicz et al. (in press, NBR)

*Bridging the gap between emotion and joint action.* | Bieńkiewicz et al. (in press, NBR)

*Bridging the gap between emotion and joint action.* | Bieńkiewicz et al. (in press, NBR)

## Annex 1

| Keywords (acting together) | Keywords (emo) | | |
|---|---|---|---|
| joint action | emotion * | | |
| acting together | | | |
| acting in unison | | | |
| moving in unison | | | |
| moving together | Keywords (clinical) | | |
| group movement | Autism | | |
| group synchronisation | ASD | | |
| action coordination | Autistic | | |
| social motor coordination | Alexithymia | | |
| action cooperation | | | |
| entrainment | | | |
| action coupling | | | |
| motor coupling | | | |
| group imitation | | | |
| motor mimicry | | | |
| action mimicry | | | |
| propagation | | | |
| synchronisation | | | |
| chameleon effect | | | |
| synchrony | | | |
| | | | REVELANT HITS (repetitions were ignored from the count) |
| PUBMED SYNTAX/ MAIN SEARCH | | | |
| (group movement [Title/Abstract] AND emotion [Title/Abstract] | | | 0/5 |
| (group synchronisation [Title/Abstract]) AND emotion [Title/Abstract] | | | 1/5 |
| (synchronisation [Title/Abstract]) AND emotion [Title/Abstract] | | | 3/24 |
| (moving in unison [Title/Abstract]) AND emotion [Title/Abstract] | | | 0 |
| (moving together [Title/Abstract]) AND emotion [Title/Abstract] | | | 4/78 |
| (acting in unison [Title/Abstract]) AND emotion [Title/Abstract] | | | 0/64 |
| (acting together [Title/Abstract]) AND emotion [Title/Abstract] | | | 2/2 |
| (action coordination [Title/Abstract]) AND emotion [Title/Abstract] | | | 1/1 |
| (social motor coordination [Title/Abstract]) AND emotion [Title/Abstract] | | | 1/39 |
| (action cooperation [Title/Abstract]) AND emotion [Title/Abstract] | | | 1/1 |
| (motor coupling [Title/Abstract]) AND emotion [Title/Abstract] | | | 0/2 |
| (action coupling [Title/Abstract]) AND emotion [Title/Abstract] | | | 2/5 |
| (motor mimicry [Title/Abstract]) AND emotion [Title/Abstract] | | | 2/11 |
| (action mimicry [Title/Abstract]) AND emotion [Title/Abstract] | | | 8/27 |
| (group imitation [Title/Abstract]) AND emotion [Title/Abstract] | | | 5/46 |
| (entrainment [Title/Abstract]) AND emotion [Title/Abstract] | | | 5/97 |
| (joint action [Title/Abstract]) AND emotion [Title/Abstract] | | | 0/5 |
| (propagation [Title/Abstract]) AND emotion [Title/Abstract] | | | 0/33 |
| (Interpersonal coordination [Title/Abstract]) AND emotion [Title/Abstract] | | | 5/25 |
| (chameleon effect [Title/Abstract]) AND emotion [Title/Abstract] | | | 2/2 |
| (group synchronization) AND emotion [Title/Abstract] | | | 3/96 |
| (motor synchronization) AND emotion [Title/Abstract] | | | 2/57 |
| ((emotion[Title/Abstract]) AND synchrony[Title/Abstract]) AND (motor[Title/Abstract] OR movement[Title/Abstract] OR motion[Title/Abstract]) | | | 11/17 |
| | | | |
| | | | |
| | | | |
| PUBMED SYNTAX/ CLINCIAL SEARCH | | | |
| AUTISTIC SPECTRUM DISORDER | | | |
| ((autism[Title/Abstract] OR autistic[Title/Abstract]) AND (joint action[Title/Abstract] OR synchronisation[Title/Abstract] OR chameleon effect[Title/Abstract] OR synchronization[Title/Abstract] OR propagation[Title/Abstract] OR action mimicry[Title/Abstract] OR motor mimicry[Title/Abstract] OR group imitation[Title/Abstract] OR motor coupling[Title/Abstract] OR action coupling[Title/Abstract] OR entrainment[Title/Abstract] OR action cooperation[Title/Abstract] OR social motor coordination[Title/Abstract] OR action coordination[Title/Abstract] OR group synchronisation[Title/Abstract] OR group movement[Title/Abstract] OR moving together[Title/Abstract] OR moving in unison[Title/Abstract] OR acting in unison[Title/Abstract] OR acting together[Title/Abstract] OR group synchronization[Title/Abstract]) | | | 40/ 441 |